\documentclass[a4paper,12pt]{article}

\usepackage{geometry,slashed}
\RequirePackage[T1]{fontenc}

\RequirePackage{graphicx}
\RequirePackage{flushend}
\RequirePackage[numbers,sort&compress]{natbib}
\RequirePackage[colorlinks,citecolor=blue,urlcolor=blue,linkcolor=blue]{hyperref}

\usepackage{booktabs}
  \usepackage{soul}
  \usepackage[usenames,dvipsnames]{xcolor}
 
 \definecolor{X575}{rgb}{0.05, 0.7, 0.05}
 
\usepackage{cancel} 
\usepackage{graphicx} 
 \usepackage{pgffor}
\usepackage{array}
\usepackage{amsmath}
 \usepackage{color}
 \usepackage{hyperref}
 \usepackage{authblk}
 \usepackage{epstopdf}
 \usepackage{enumerate}
 \usepackage{multirow}
 \usepackage{siunitx}
 \usepackage{float}
 \usepackage{comment}

  \usepackage{listings}
  \usepackage{subfigure}

 \newcommand{\tllj}{t\ell^+\ell^-j}

 \newcommand{\MSbar}{{\rm \overline{MS}}}
  \newcommand{\TO}{\rightarrow}
  
   \newcommand{\taj}{t\gamma j}
   \newcommand{\tta}{t \bar t\gamma }
   \newcommand{\ttaa}{t \bar t\gamma \gamma }
      \newcommand{\tadlep}{t \rightarrow b \ell^{+}\nu_{\ell} \gamma}
\newcommand{\tadhad}{t \rightarrow b jj \gamma}

\newcommand{\gev}{\textrm{GeV}}

\newcommand{\mt}{m_t}
\newcommand{\mz}{m_Z}
\newcommand{\mw}{m_W}

\newcommand{\mglong}{{\sc\small MadGraph5\_aMC@NLO}}

\newcommand{\gmu}{G_\mu}

\newcommand\mf{{\sc\small MadFKS}}
\newcommand\ml{{\sc\small MadLoop}}
\newcommand\ct{{\sc\small CutTools}}
\newcommand\nin{{\sc\small Ninja}}
\newcommand\coll{{\sc\small Collier}}

\def\alphas{\alpha_s}
\def\alphaz{\alpha(0)}
\def\alphaGmu{\alpha_{G_\mu}}

\def\TO{{\longrightarrow}}
\def\ord{\mathcal{O}}
\def\alphas{\alpha_s}
\def\LO{{\rm LO}}
\def\NLO{{\rm NLO}}
\def\NLOQCD{{\rm NLO_{QCD}}}
\def\LOQCD{{\rm LO_{QCD}}}
\def\NLOQCDt{\NLO_{{\rm QCD},t{\rm -ch.}}}
\def\NLOQCDEW{\NLO_{{\rm QCD+EW}}}

\def\ttgamma{t \bar t \gamma}

\title{Automated EW corrections with isolated photons: \\
 $t \bar t \gamma$, $t \bar t \gamma\gamma$ and $t \gamma j$ as case studies}

\author[1]{Davide Pagani\thanks{davide.pagani@desy.de}}
\author[2]{Hua-Sheng Shao\thanks{huasheng.shao@lpthe.jussieu.fr}}
\author[3]{Ioannis Tsinikos\thanks{ioannis.tsinikos@thep.lu.se}}
\author[4]{Marco Zaro\thanks{marco.zaro@mi.infn.it}}

\affil[1]{\small DESY, Theory Group, Notkestrasse 85, 22607 Hamburg, Germany}
\affil[2]{\small Laboratoire de Physique Th\'eorique et Hautes Energies (LPTHE), UMR 7589, Sorbonne Universit\'e et CNRS, 4 place Jussieu, 75252 Paris Cedex 05, France}
\affil[3]{\small Theoretical Particle Physics, Department of Astronomy and Theoretical Physics, Lund University, S\"olvegatan 14A, SE-223 62 Lund, Sweden}
\affil[4]{\small TIF Lab, Dipartimento di Fisica, Universit\`a degli Studi di Milano \& INFN, Sezione di Milano, Via Celoria 16, 20133 Milano, Italy}

\begin{document}
 
  \date{}
 
 \maketitle
 
  \vspace*{-10cm}
  {
      {{\color{blue}DESY 21-083}  \hfill
      \color{blue}{LU-TP 21-19}
      \hfill \color{blue}{TIF-UNIMI-2021-7}}
  }
  \vspace*{9cm}

\begin{abstract}
    In this work we compute for the first time the so-called Complete-NLO predictions  for top-quark pair hadroproduction in association with at least one isolated photon ($ t \bar t \gamma$). We also compute NLO QCD+EW  predictions for the similar case with at least two isolated photons  ($ t \bar t \gamma\gamma$) and for single-top hadroproduction in association with at least one  isolated photon. In addition, we complement our results with NLO QCD+EW predictions of the hadronic and leptonic decays of top-quark including an isolated photon. All these results have been obtained in a completely automated approach, by extending the capabilities of the {\sc\small MadGraph5\_aMC@NLO} framework and  enabling the Complete-NLO predictions for  processes with isolated photons in the final state. We  discuss the technical details of the implementation, which involves a mixed EW renormalisation scheme for such processes.  
  \end{abstract}

 \section{Introduction}\label{sec:intro}
 
 After the first ten years of operation of the Large Hadron Collider (LHC), our knowledge of the fundamental interactions of elementary particles has been tremendously improved. At the LHC, the long-searched Higgs boson has been observed~\cite{Aad:2012tfa,Chatrchyan:2012ufa}, and its properties have also been studied in details and found to be compatible with those predicted by the Standard Model (SM) \cite{Aad:2019mbh}. Moreover, the SM itself, our current best understanding of elementary particles and fundamental interactions, has been stress-tested not only for what concerns the Higgs sector, but in almost all other aspects, {\it e.g.}, electroweak (EW) interactions, QCD dynamics and flavour physics. So far, no clear and unambiguous sign of beyond-the-SM (BSM) physics has been found at colliders, but the BSM search programme at the LHC is still only at the initial phase, since $20$ times more data will be collected in the coming years, large part of it during the High-Luminosity (HL) runs \cite{Azzi:2019yne,Cepeda:2019klc,CidVidal:2018eel,Cerri:2018ypt,Citron:2018lsq,Chapon:2020heu}.
 
The success of this ambitious research programme relies on the ability of providing precise and reliable SM predictions. For this reason, in the past years, a plethora of new calculations and techniques have appeared in the literature, aiming to improve SM (but also BSM) predictions. On the one hand, a lot of efforts have been put in improving the calculations of purely QCD radiative corrections, going from Next-to-Leading-Order (NLO) to Next-to-NLO (NNLO) or even Next-to-NNLO ($\rm N^3LO$) predictions
and in parallel improving the resummation of large logarithms appearing at fixed order. On the other hand, a lot of work has been done for the calculations of NLO QCD and EW corrections for processes with high-multiplicity final states. To this purpose, NLO QCD and EW corrections have been implemented in Monte Carlo generators and, at different levels in the different frameworks, they have been even automated \cite{ Kallweit:2014xda, Frixione:2015zaa, Chiesa:2015mya, Frederix:2016ost, Biedermann:2017yoi,  Frederix:2018nkq}.   

As explained in detail in Ref.~\cite{Frederix:2018nkq},  the automation of NLO corrections of QCD and EW origins has been implemented in the {\mglong} framework \cite{Alwall:2014hca}. Thanks to this, not only NLO QCD+EW corrections for many new hadroproduction processes have become available \cite{ Frixione:2015zaa, Pagani:2016caq, Frederix:2016ost, Czakon:2017wor, Frederix:2017wme, Frederix:2018nkq, Broggio:2019ewu, Frederix:2019ubd, Pagani:2020rsg, Pagani:2020mov}, but also the subleading NLO effects can be systematically assessed. In other words, all perturbative orders arising from tree-level diagrams and their interference with their one-loop counterparts in the SM can be computed in the automated means. This has lead to the discovery that, in some cases, the subleading NLO corrections (those beyond the standard NLO QCD $\ord(\alphas)$ and NLO EW $\ord(\alpha)$ corrections)  can be much larger than their naive estimates based on the power counting of $\alphas$ and $\alpha$~\cite{ Biedermann:2016yds, Frederix:2017wme, Pagani:2020rsg}. In particular, this has been observed in the context of top-quark physics for the production of a top-quark pair in association with a $W$ boson ($t \bar t W$) \cite{Frederix:2017wme, Broggio:2019ewu, Kulesza:2020nfh, Frederix:2020jzp, Cordero:2021iau, Denner:2021hqi} and for the production of four top quarks ($t \bar t t \bar t $) \cite{Frederix:2017wme}. 

While the theoretical framework of Ref.~\cite{Frederix:2018nkq} is general and applicable to any possible final state, the formalism regarding fragmentation functions has not been implemented yet in the {\mglong} code. In particular, this formalism is necessary when measurable quantities are
not defined only by means of massive particles and/or after a jet-clustering(-like) algorithm. Consequently, so far, one of the main limitations of the code has been the impossibility of calculating NLO EW corrections and Complete-NLO predictions for  processes with tagged photons. 

The aim of this paper is twofold. First, we want amend to the aforementioned limitations, allowing NLO EW and Complete-NLO calculations for processes involving photons that are tagged by applying an isolation algorithm \cite{Frixione:1998jh}. Second, we want to exploit the new capabilities of the code for computing NLO EW and Complete-NLO predictions involving both photons and top quarks. Since, as previously mentioned, unexpected large NLO EW radiative effects have been observed in similar processes involving top quarks and massive vector bosons, it is natural to check the same thing for processes involving top quarks and isolated photons.

The automation of EW corrections involving isolated photons has been achieved by using the so-called $\alphaz$ renormalisation scheme in the {\mglong} framework. Its implementation is in fact performed via a mixed-scheme approach \cite{Andersen:2014efa, Denner:2019vbn}, which is based on the idea that $\alpha$ should be renormalised in the $\alphaz$-scheme only for (final-state) isolated photons, while other EW interactions should be renormalised in the $\alpha(\mz)$ or $G_\mu$-scheme. We acknowledge that a similar procedure has already been implemented in a matrix-element provider, namely {\sc\small OpenLoops2}~\cite{Buccioni:2019sur}.  However, this is the first time that this approach has been pursued in the context of a fully-fledged automation of NLO EW corrections, as done in {\mglong}.  We provide technical details and make necessary clarifications on this subject and we also discuss the issues concerning the choice of the numerical value of $\alpha$ in the $\ord(\alpha)$ corrections.

We exploit the new implementation for computing precise predictions for top-quark pair hadroproduction in association with at least one isolated photon ($ t \bar t \gamma$),  with at least two isolated photons  ($ t \bar t \gamma\gamma$) and for single-top hadroproduction in association with at least one  isolated photon ($\taj$). We also complement our results with the case of the hadronic ($\tadhad$) and leptonic ($\tadlep$) decays of a top quark including an isolated photon.
The measurement and the analysis of this class of processes is crucial for testing the interaction of top quarks and photons and for determining top-quark properties ({\it e.g.}~its electric charge~\cite{Melnikov:2011ta} and electromagnetic dipole moments~\cite{Fael:2013ira}), and possibly detect BSM deviations ({\it e.g.}~due to the flavour-changing interactions~\cite{Durieux:2014xla} or anomalous top-photon couplings~\cite{Etesami:2016rwu,Etesami:2018mqk}). Studies in this direction, especially in the context of the SM effective field theory (SMEFT) \cite{Bylund:2016phk, Schulze:2016qas, Englert:2017dev, Degrande:2018fog, Durieux:2019rbz,Brivio:2019ius, Maltoni:2019aot}, have been performed and revealed their special roles.\footnote{We also reckon that the possibility to merge processes with different photon multiplicities has been proposed in Ref.~\cite{BuarqueFranzosi:2020baz}, in order to simulate backgrounds to BSM searches.}

It is obvious that precise SM predictions for this class of processes must be available  in the first place in order to probe possibly tiny  BSM effects. Thus, higher-order perturbative corrections of both QCD and EW origins have to be included in the LHC phenomenological applications. To this purpose, we carry out the first Complete-NLO computation of $ t \bar t \gamma$ hadroproduction. We also compute for the first time NLO QCD+EW  predictions for $ t \bar t \gamma\gamma$ and $\taj$ production processes. For the last process, we also consider the different flavour-scheme dependence, following the approach presented in Ref.~\cite{Pagani:2020mov}. Finally,  we apply our framework to calculate NLO QCD+EW corrections, for the first time, for the hadronic and leptonic  top-quark decays including an isolated photon.

The structure of the paper is as follows.  In Sec.~\ref{sec:implementation}, we describe the automation of EW corrections and more in general of Complete-NLO predictions in the {\mglong} framework for isolated photons. We start discussing the notation and the general approach to the problem in Sec.~\ref{sec:notationandmore}, where we also show the syntax that should be used in order to run the code. Then, in Sec.~\ref{sec:technical} we discuss the technical details.
In Sec.~\ref{sec:processes}, we present results at the inclusive and differential level for the processes mentioned before. Finally, in  Sec.~\ref{sec:conclusions}, we draw our conclusions.

\section{Automation of EW corrections with isolated photons}
 \label{sec:implementation}
 
The automation of EW corrections and more in general of Complete-NLO predictions in the {\mglong} framework has been presented and discussed in details in Ref.~\cite{Frederix:2018nkq}. Therein, the theoretical set-up has been based on the use of renormalisation schemes where infrared (IR) divergencies of renormalised one-loop amplitudes are  $\MSbar$-like,\footnote{The ultraviolet (UV) poles of a given counterterm are always identical in different renormalisation schemes, while the finite parts are different. The IR poles are exactly zero in the {$\MSbar$} scheme and therefore we will call those schemes with this feature as ``{$\MSbar$}-like".}  such as the $\MSbar$ scheme itself but also, for the EW sector, the more commonly used $\alpha(\mz)$ and especially  $G_\mu$-scheme. Using this class of renormalisation schemes, if there is a massless particle $p_i$ that can spilt into two massless particles $p_j$ and $p_k$ via an EW splitting $p_i\longrightarrow p_j p_k$, an NLO EW calculation cannot be straightforwardly carried out for a final state including $p_i$ as a physical object. The calculation would be IR divergent. In the SM, photons and all the charged massless fermions, can precisely split via QED interactions into two massless particles.
IR safety can be in general achieved via two different solutions: clustering massless particles into fully-democratic jets\footnote{In many occasions it is sufficient to use simpler definitions, such as, {\it e.g.}, dressed leptons. } or using fragmentation functions. The definition and the scope of the former has been extensively explored in Ref.~\cite{Frederix:2016ost}, while the description of the necessary theoretical framework for the implementation of fragmentation functions in {\mglong} has be presented in Ref.~\cite{Frederix:2018nkq}. The usage of fragmentation function in the context of an NLO EW calculation has also been exploited in Ref.~\cite{Denner:2019zfp}.

On the other hand, for the case of photons in the final state, a very well known and (probably much) simpler solution than the usage of fragmentation functions exists: performing perturbative calculations in the $\alphaz$-scheme. In fact, concerning the purely QED part of NLO EW corrections,  besides effects that are formally beyond NLO and related to the fragmentation-function evolution,  a calculation performed in an $\MSbar$-like renormalisation scheme employing the photon fragmentation function and a calculation performed in the $\alphaz$-scheme with isolated-photons lead to the same result, as shown, {\it e.g.}, in Ref.~\cite{Frederix:2016ost} and discussed also in Sec.~\ref{sec:technical}. In the following context, we will in general understand that the Frixione isolation algorithm \cite{Frixione:1998jh} is employed for isolating the photon.  

 In the following of this section we describe the modifications and extensions, w.r.t.~the theoretical framework  in Ref.~\cite{Frederix:2018nkq}, that we have implemented in {\mglong}. Via these new features, NLO EW corrections are enabled for any process involving isolated photons in the final state. Moreover,  Complete-NLO predictions can be calculated for any process, besides some cases involving simultaneously both isolated photons and jets. We will return to it later, at the end of  Sec.~\ref{sec:fks},  on this limitation and explain the reason behind it.

There are three main improvements w.r.t~the framework described in Ref.~\cite{Frederix:2018nkq}. They concern the following three aspects:
\begin{itemize}
\item the renormalisation conditions,
\item the FKS counter-terms,
\item the photon isolation together with democratic jets.
\end {itemize}

Before describing the technical part of these aspects in Sec.~\ref{sec:technical}, we define the necessary notations and describe the general approach to the problem in Sec.~\ref{sec:notationandmore}. Besides, we also show the syntax that should be used in order to run the code.\footnote{The new features of the {\mglong} framework described in this work will become public in a future release of the code.}

\subsection{Notation, syntax and calculation set-up}
\label{sec:notationandmore}

\subsubsection{Notation}
\label{sec:notation}

Adopting the notations already used in Refs.~\cite{Frixione:2014qaa, Frixione:2015zaa, Pagani:2016caq, Frederix:2016ost, Czakon:2017wor, Frederix:2017wme, Frederix:2018nkq, Broggio:2019ewu, Frederix:2019ubd,Pagani:2020rsg, Pagani:2020mov}
the different contributions from the expansion in powers of $\alpha_s$ and $\alpha$ of  any differential or inclusive cross section $\Sigma$ at LO ($\Sigma^{}_{\LO}$) and at NLO ($\Sigma^{}_{\NLO}$) can be denoted as:
\begin{align}
\Sigma^{}_{\LO}(\alpha_s,\alpha) &= \Sigma^{}_{\LO_1} + \ldots  + \Sigma^{}_{\LO_k}\, , \\
 \Sigma^{}_{\NLO}(\alpha_s,\alpha) &=  \Sigma^{}_{\NLO_1} + \ldots  + \Sigma^{}_{\NLO_{k+1}} , \label{eq:blobs_NLO_general} 
\end{align}
 where  $k\ge1$ and the specific value of $k$ is process dependent. Each $\Sigma^{}_{\LO_i}$ denotes a different $\alphas^n \alpha^m$ perturbative order stemming at LO, {\it i.e.}, from Born diagrams only. In a given process, both the values of $n$ and $m$ are different for each $\Sigma^{}_{\LO_i}$, but the sum $n+m$ is fixed.  If $\Sigma^{}_{\LO_i}\propto \alphas^n \alpha^m$ then $\Sigma^{}_{\LO_{i+1}}\propto \alphas^{n-1} \alpha^{m+1}$, $\Sigma^{}_{\NLO_{i}}\propto \alphas^{n+1} \alpha^{m}$ and $\Sigma^{}_{\NLO_{i+1}}\propto \alphas^{n} \alpha^{m+1}$, where each $\Sigma^{}_{\NLO_i}$ denotes a different NLO perturbative order stemming from the interference between Born and one-loop diagrams. 
  
The quantity denoted as $\Sigma^{}_{\LO_1}$  is what is commonly referred as $\LO$ in the literature, while here ``LO'' denotes the sum of all the possible $\Sigma^{}_{\LO_i}$.  We will also use the standard notations ``$\NLOQCD$'' and ``$\NLOQCDEW$'' for the quantities  $\Sigma^{}_{\LO_1}+\Sigma^{}_{\NLO_1}$ and $\Sigma^{}_{\LO_1}+\Sigma^{}_{\NLO_1}+\Sigma^{}_{\NLO_2}$, respectively.  The $\Sigma^{}_{\NLO_1}$ and $\Sigma^{}_{\NLO_2}$ terms are in other words the NLO QCD and NLO EW corrections, respectively. We will also use in general the alias ``$\rm (N)LO_i$'' in order to indicate the quantity $\Sigma^{}_{\rm (N)LO_i}$.  The set of all the possible contributions of $\ord(\alpha_s^n \alpha^m)$ at LO and NLO is what is denoted as ``Complete-NLO''.

\subsubsection{Calculation set-up with isolated photons}
\label{sec:set-up}

Following the strategy described in Refs.~\cite{Andersen:2014efa, Denner:2019vbn}, for processes including  isolated photons in the final state we perform the renormalisation of EW corrections in a mixed scheme.  Any $\NLO_{i}$ term with $i\ge 2$ for a process including isolated photons involves the renormalisation of EW interactions, or equivalently of the powers of $\alpha$  that are present in the $\LO_{i-1}$. The standard case of NLO EW corrections correspond to $i=2$. For a general process involving $n_\gamma$ isolated photons,
\begin{equation}
p p \TO n_{\gamma} \gamma_{\rm iso} + X\, , \label{eq:genproc}
\end{equation}
if $\Sigma^{}_{\LO_i}\propto \alphas^n \alpha^m$, then $m\ge n_\gamma$ and there are $n_\gamma$ powers of $\alpha$ related to the vertices with final-state external photons and $m-n_\gamma$ ones of a different kind. While $\MSbar$-like schemes like the $\alpha(\mz)$ and especially the $G_\mu$-scheme are in general superior to the $\alphaz$-scheme for the calculation of the EW corrections, in the case of final-state legs associated to isolated photons it is the opposite (see {\it e.g.}~the aforementioned Refs.~\cite{Andersen:2014efa, Denner:2019vbn} for more details about it). Actually, in modern calculations where light-fermion masses are set equal to zero, this choice of scheme is not only superior but also necessary to achieve IR safety. We will show this in more details in Sec.~\ref{sec:ren}. Therefore, we renormalise $n_\gamma$ powers of $\alpha$ in the $\alphaz$-scheme, while $m-n_\gamma$ powers in an other $\MSbar$-like scheme, which in the rest of the article will be, if not differently specified, the $G_\mu$-scheme.\footnote{The same procedure described in the following could be framed also with the $\alpha(\mz)$-scheme in place of the $G_\mu$-scheme, although the latter should be in general preferred  rather than the former. } It is worth to stress that the $\alphaz$-scheme should {\it not} be adopted for initial-state photons~\cite{Harland-Lang:2016lhw, Kallweit:2017khh}.  
The usage of the $\alphaz$-scheme also implies that the standard procedure for the generation of real-radiation diagrams described in Ref.~\cite{Frederix:2018nkq} has to be modified. In particular, since the final-state photon in a diagram coincides with a physical object, the isolated-photon, final-state QED $\gamma\TO f \bar f $ splittings , where $f$ is a charged massless fermion, should be vetoed. In other words, for the process in eq.~\eqref{eq:genproc} the real radiation diagrams with  final state  $(n_{\gamma}-1) \gamma_{\rm iso} + f \bar f + X$ leading to the same perturbative order of $\NLO_{i+1}$ should not be taken into account in the calculation. For this reason, similarly to the virtual contribution, the divergencies arising from real radiation are different than in an $\MSbar$-like scheme. Thus, the definition of FKS counterterms should be amended too (see details in Sec.~\ref{sec:fks}).

\medskip

When performing a calculation, the differences between two specific renormalisation schemes do not only consist of the different renormalisation counter terms. Indeed, also the numerical values that should  be used for the input parameters are different. In the case of the $\alphaz$-scheme and $G_\mu$-scheme, the numerical values for the QED fine structure constant $\alpha$ are different, namely  $\alpha=\alphaz$ and $\alpha=\alphaGmu$ with
\begin{equation}
\alphaz\simeq\frac{1}{137}\, \qquad \alphaGmu=\frac{\sqrt{2}G_\mu\mw^2}{\pi}\left(1-\frac{\mw^2}{\mz^2}\right)\simeq \frac{1}{132}.
\end {equation}
Since for a process with $n_\gamma$ isolated photons in the final state and with $\Sigma^{}_{\LO_i}\propto \alphas^n \alpha^m$ we renormalise $n_\gamma$ powers of $\alpha$ in the $\alphaz$-scheme and  $m-n_\gamma$ powers in the $G_\mu$-scheme, 
we consistently set the input parameters according to the rule
\begin{equation}
\Sigma^{}_{\LO_i}\propto \alphas^n \alpha^m \Longrightarrow \Sigma^{}_{\LO_i}\propto \alphas^n (\alphaGmu^{m-n_\gamma}\alphaz^{n_\gamma}) \,. \label{eq:LOa0Gmu}
\end{equation}

At NLO accuracy, for what concerns the input parameters, it is instead necessary  to differentiate  two separate cases on the basis of the power of $\alphas$ in $\Sigma^{}_{\LO_i}$:  $n>0$, which allows for the presence of $\Sigma^{}_{\LO_{i+1}}$, and  $n=0$, which in the SM model implies that $\Sigma^{}_{\LO_{i+1}}$ is not present.
If $n>0$, at variance with the LO case, in $\Sigma^{}_{\NLO_{i+1}}$ there is in general no freedom of choice in the numerical value of the additional power of $\alpha$ without spoiling the cancellation of  UV and/or IR divergencies. The numerical value of the additional power of $\alpha$, which we denote as $\bar \alpha$, has to be set equal to $\alphaGmu$, {\it i.e.},
\begin{equation}
\Sigma^{}_{\LO_i}\propto \alphas^n (\alphaGmu^{m-n_\gamma}\alphaz^{n_\gamma}) \Longrightarrow\, 
\Sigma^{}_{\NLO_{i+1}}\propto \alphas^n \bar{\alpha}(\alphaGmu^{m-n_\gamma}\alphaz^{n_\gamma})\,\qquad{\rm with} \qquad \bar{\alpha}=\alphaGmu. \label{eq:NLO_rescale_fixed}
\end{equation}
Indeed, the $\alphas$ and $\alpha$ expansion of $\Sigma^{}_{\LO}$ is actually an expansion in $\alphas$ and $\alphaGmu$, as can be seen in \eqref{eq:LOa0Gmu}. Since ${\Sigma^{}_{\LO_{i}}}/{\Sigma^{}_{\LO_{i+1}}}\propto \alphas/\alphaGmu$ and
\begin{equation}
\ord({\Sigma^{}_{\NLO_{i+1}}})= \ord({\Sigma^{}_{\LO_{i}}})\times\bar \alpha = \ord({\Sigma^{}_{\LO_{i+1}}})\times \alphas\, , \label{eq:abar_const}
\end{equation}
the only choice of $\bar \alpha$ that in general preserves the exact cancellation of both UV and IR divergencies is $\bar \alpha=\alphaGmu$.

If instead $n=0$, namely no QCD interactions in $\Sigma^{}_{\LO_i}$,  the quantity $\Sigma^{}_{\LO_{i+1}}$ does not exist and therefore eq.~\eqref{eq:abar_const} does not imply that the relation  $\bar \alpha=\alphaGmu$ must be true in order to preserve the exact cancellation of both UV and IR divergencies. In other words, 
\begin{equation}
\Sigma^{}_{\LO_i}\propto (\alphaGmu^{m-n_\gamma}\alphaz^{n_\gamma}) \Longrightarrow\, 
\Sigma^{}_{\NLO_{i+1}}\propto  \bar{\alpha}(\alphaGmu^{m-n_\gamma}\alphaz^{n_\gamma})\,\qquad{\rm with} \qquad \alphaz\le\bar{\alpha}\le\alphaGmu\, \label{eq:NLO_rescale}
\end{equation}
and the choice of the numerical value of $\bar \alpha$ is in principle arbitrary, being an NNLO $\ord(\alpha^2)$ effect w.r.t.~$\Sigma^{}_{\LO_i}$. 
The same argument could be repeated for any other $\MSbar$-like scheme in the place of the $G_\mu$-scheme.

On the other hand, also for $n = 0$, the choice $\bar \alpha=\alphaGmu$ should be in general preferred and regarded as superior than $\bar \alpha=\alphaz$. Indeed, NLO corrections do not contain any additional isolated photon, since the additional photons appearing via real radiation are unresolved.\footnote{One can understand this also from the fact that, {\it e.g.}, NNLO $\ord(\alpha^2)$ corrections would involve both additional single and double real radiation of light particles. In the former class, one-photon emissions at one-loop would be present. In the latter class, tree-level one-photon emission with further $\gamma\TO f \bar f$ splitting would be also present and {\it not} vetoed. Therefore the single emission should be parametrised by $\bar \alpha=\alphaGmu$ rather than $\bar \alpha=\alphaz$. } The only case in which $\bar \alpha=\alphaz$ is preferable is when $\Sigma^{}_{\NLO_{i+1}}$ predictions are used for observables involving $n_\gamma+1$ isolated photons, but for this case a calculation involving $n_\gamma+1$ isolated photons already at the tree level, therefore proportional to $\alphaz^{n_\gamma +1}$ according to \eqref{eq:LOa0Gmu}, should be preferred. Still, it is important to note that the choice of the value of $\bar \alpha$ , as already said, formally affects NNLO $\ord(\alpha^2)$ corrections and  especially its impact in \eqref{eq:NLO_rescale} is typically  at the permille and more often at sub-permille level on the prediction of an observable. Indeed, if we consider for instance the NLO EW corrections ($\NLO_2$), the impact of this choice w.r.t.~the dominant LO, the $\LO_1$, is of the order $(\NLO_2/\LO_1)\Delta \bar{\alpha}$, with  $\Delta\bar{\alpha}\equiv(\alphaGmu-\alphaz)/\bar \alpha \simeq 0.04$, with the quantity $(\NLO_2/\LO_1)$ being also at the percent level. 

Finally, we notice something quite counterintuitive that is a consequence of setting $\bar \alpha=\alphaGmu$. Even with $n_\gamma=m$ (all EW interactions being associated to vertices involving isolated photons), the mixed scheme is not fully equivalent to the pure $\alphaz$ scheme precisely because $\bar \alpha=\alphaGmu$. Not only, for the same reasons already explained before, the mixed scheme is also in this case superior to the pure $\alphaz$ scheme. Moreover, if on top of that $n>0$, although all the EW final-state objects are isolated photons,  the condition $\bar \alpha=\alphaGmu$ is still in general necessary for IR-safety and UV-finiteness, due to eq.~\eqref{eq:abar_const}.\footnote{
In principle, one could  set $\bar \alpha=\alphaz$, but it would be necessary to alter eq.~\eqref{eq:LOa0Gmu} by using only $\alphaz$ as the input parameter for all powers of $\alpha$, which is clearly not a good choice. On the other hand, this procedure would lead for processes with $n_\gamma=m$ to the exactly the same results obtainable with the pure $\alphaz$ scheme, also with  $n>0$.}

We want to stress that all this discussion on $\bar \alpha$ is particularly relevant in the context of a fully-fledged automation. If analytical expressions are available and/or is possible to separate subsets of diagrams or contributions that are separately IR and UV finite, further sophistications employing separate and optimised input values for $\alpha$, as well as other parameters, can be performed. On the other hand, with the previous discussion we want to emphasise that in an automated calculation the only safe procedure is setting a common value $\bar \alpha$ for all the contributions to the $\Sigma^{}_{\NLO_{i+1}}$, and especially to show to which value $\bar \alpha$ should or can be set.
\subsubsection{Generation  syntax}

After having specified the notation and the general aspects of the theoretical set-up, we now illustrate the commands that have to be used in the {\mglong} framework in order to calculate EW corrections for processes involving isolated photons. We have introduced the notation {\tt!a!} for an isolated photon in the generation syntax of the framework. Let us present a few concrete examples used in this paper.

First of all, the correct model have to be imported. One can choose either\footnote{For processes with only isolated photons in the final state, also the pure $\alphaz$ scheme {\tt loop\_qcd\_qed\_sm\_a0} can be loaded.}
\begin{verbatim}
import model loop_qcd_qed_sm_a0-Gmu
\end{verbatim}
or
\begin{verbatim}
import model loop_qcd_qed_sm_Gmu-a0
\end{verbatim}
Both of them work accordingly to  the mixed scheme described in the previous section. However, the former corresponds to the choice $\bar \alpha = \alphaz$ in \eqref{eq:NLO_rescale}, while the latter to the choice $\bar \alpha = \alphaGmu$. As explained, the second option is the only one that, starting from \eqref{eq:LOa0Gmu}, in general satisfies the necessary condition from eq.~\eqref{eq:abar_const}. Moreover, as also already explained, is superior from a formal point of view and should be in general preferred. In the results presented in Sec.~\ref{sec:processes} we will use this option, unless differently specified. 
Then, if we want to calculate a single top associated hadroproduction process at NLO QCD+EW accuracy we use the following syntax:
\begin{verbatim}
generate p p > t j !a! [QCD QED]
\end{verbatim}
where we have taken an example studied in this paper. If one is only interested in NLO QCD or NLO EW corrections, the  {\tt QED} or {\tt QCD} flag in the squared bracket should be respectively omitted.
If we are intending to calculate the Complete-NLO predictions to the $\ttgamma$ hadroproduction process, one has to use 
\begin{verbatim}
generate p p > t t~ !a! QCD^2=100 QED^2=100 [QCD QED]
\end{verbatim}
to generate the process. Similarly, if we restrict to only $\LO_2 \propto \alphas \alpha^2$ and $\NLO_3 \propto \alpha_s \alpha^3$ terms, one has to type
\begin{verbatim}
generate p p > t t~ !a! QCD^2=2 QED^2=4 [QED]
\end{verbatim}
In general, in order to select a LO contribution proportional to $ \propto \alphas^n \alpha^m$ the tag {\tt QCD\^{}2} should be set to $2 n$ and the tag {\tt QED\^{}2} should be set to $2 m$.\footnote{For the reader that is used to the {\mglong} code, we want to stress that at variance with previous versions of the code, from the version 3.1 onwards the syntax {\tt QCD\^{}2=2n} and {\tt QCD=n} are {\it not} equivalent anymore, and similarly with {\tt QED}. The second should be now avoided by the non-expert user. See also the webpage \url{http://amcatnlo.web.cern.ch/amcatnlo/co.htm}.} We want to stress that the most important point at the generation level is the usage of {\tt!a!} for isolated photons, which is not equivalent to the simple {\tt a} (non-isolated photon). Only the former prevents the $\gamma\TO f \bar f$ splitting, which is necessary for the consistency in the NLO calculation with isolated photons. 

Before providing the technical details we want to mention that we have cross-checked results obtained in a completely automated way against  calculations already present in the literature for the hadroproduction of the $
\ell^+\ell^-/\bar\nu\nu+\gamma$ \cite{Denner:2015fca}, $\gamma\gamma$ \cite{Bierweiler:2013dja} and $\gamma\gamma\gamma$ \cite{Greiner:2017mft} final states, finding perfect agreements.

\subsection{Technical details}
\label{sec:technical}

We now provide the technical details of the calculation set-up outlined in Sec.~\ref{sec:set-up}.
\subsubsection{Renormalisation and its implementation}
\label{sec:ren}
The renormalisation of UV divergent amplitudes involves the transition from bare  to renormalised quantities, which in the EW sector involves $e\rightarrow e (1+\delta Z_e)$ or equivalently $\alpha\rightarrow \alpha (1+2\delta Z_e)$.
The $\alphaz$-scheme corresponds to the definition
\begin{equation}
\delta Z_e|_{\alpha(0)}=-\frac{1}{2} \delta Z_{AA}-\frac{s_W}{c_W}\frac{1}{2} \delta Z_{ZA} \, ,\label{deSM}
\end{equation}
where $\delta Z_{AA}$ is the wave-function  renormalisation constant of the photon, with $\delta Z_{AA}=-\Pi^{AA}(0)$, {\it i.e.}, the vacuum polarisation at virtuality equal to zero. Similarly, $\delta Z_{ZA}$ is the non-diagonal entry of the $(A,Z)$ wave-function  renormalisation. The terms $s_W$ and $c_W$ are the sine and cosine of the Weinberg angle, respectively.
With this definition, if we would retain the masses $m_f$ of all the charged fermions in the SM, 
\begin{equation}
\delta Z_e|_{\alpha(0)}=  \frac{1}{2}\sum_f \frac{\alpha}{3 \pi}Q^2_f N^f_C (\Delta +\log(\mu^2/m_f^2))+\ldots\, ,
\label{deltaeexplicitSM}
\end{equation} where in $d=4-2\epsilon$ dimensions 
$\Delta=1/\epsilon-\gamma_E+\log(4\pi)$ and $\mu$ is the regularisation scale.
 $Q_f$ is the charge of the fermion and $N^f_C$ is the corresponding colour factor ($N^f_C=1$ for the leptons, $N^f_C=3$ for the quarks).
UV divergencies correspond to the $\Delta$  term, while the logarithms in  eq.~\eqref{deltaeexplicitSM} corresponds to IR divergencies in the massless limit $m_f=0$, which would lead to extra $1/\epsilon$ poles.  These are precisely the poles that would not be present in an $\MSbar$-like scheme. The symbol ``$\ldots$'' stands for all the remaining terms of $\delta Z_e|_{\alpha(0)}$: weak contributions and  QED terms that are neither $1/\epsilon$ poles nor logarithms involving  $m_f$.

In a process like the one in \eqref{eq:genproc}, the external final-state photons are on-shell, exactly as in the  kinematic configuration for which $\alphaz$ is defined and eq.~\eqref{deSM} is derived. Therefore, $\delta Z_e|_{\alpha(0)}$ cancels exactly the (UV and IR) poles emerging from one-loop corrections connected to the vertex where the external photon is attached to the full process. On the contrary, in an  $\MSbar$-like scheme, the UV poles would be canceled but the IR ones would not; only by combining the renormalised one-loop contribution with the integrated real-emission $(n_{\gamma}-1) \gamma_{\rm iso} + f \bar f + X$ final state the IR divergencies would be canceled. For all the other vertices in the processes, the situation is opposite. A renormalisation in an $\MSbar$-like scheme leads to the cancellation of UV divergencies, but in the $\alphaz$-scheme it introduces also a term of order $\alpha\log(Q^2/m_f^2)$, where $Q$ is the scale associated with the specific interaction vertex. With massive fermions, this is a sign of a wrong choice of the renormalisation scheme, leading to artificially enhanced corrections at large energies. With massless fermions, the calculation is simply IR divergent. To overcome these problems, our solution is precisely the mixed-scheme described in Sec.~\ref{sec:notation}.

The procedure for automating the implementation of mixed renormalisation is the following. First, we start with the case of \eqref{eq:NLO_rescale} and then we move to the case of \eqref{eq:NLO_rescale_fixed}, which in this context is a simplified version of \eqref{eq:NLO_rescale}. 
Let us consider process with $n_\gamma$ isolated photons in the final state and with $\Sigma^{}_{\LO_i}\propto  \alpha^m$, therefore $n_\gamma$ powers of $\alpha$ in the $\alphaz$-scheme and  $m-n_\gamma$ powers in the $G_\mu$-scheme. First, one has to perform the calculation in either the $\alphaz$-scheme or the $G_\mu$-scheme. After that, one has to either add the quantity $(m-n_\gamma)\,\Delta_{G_\mu,\alphaz}\,\Sigma^{}_{\LO_i}$ to the virtual contribution or subtract   $n_\gamma \,\Delta_{G_\mu,\alphaz}\, \Sigma^{}_{\LO_i} $ to it, respectively, where
\begin{equation}
\Delta_{G_\mu,\alphaz}\equiv\delta\alphaGmu-\delta\alphaz=2\alpha(\delta Z_e|_{G_\mu}-\delta Z_e|_{\alpha(0)})\,.
\end{equation}
After that, one can rescale both the $\LO_i$ and $\NLO_i$ contributions in order to achieve the prescription in \eqref{eq:NLO_rescale}, namely, multiplying both results by either $(\alphaGmu/\alphaz)^{m-n_\gamma}$ or $(\alphaz/\alphaGmu)^{n_\gamma}$, respectively.  We notice that while in the latter case the rescaling factor is only depending on the number of isolated photons, in the former it also depends on the considered QED perturbative order $m$.
 The choice of which scheme to start with corresponds to the choice of the value of $\bar \alpha$ in \eqref{eq:NLO_rescale}, either $\bar \alpha=\alphaz$ or $\bar \alpha=\alphaGmu$, respectively, and in turn on which model is imported when performing the calculation in {\mglong}, {\tt loop\_qcd\_qed\_sm\_a0-Gmu} or {\tt loop\_qcd\_qed\_sm\_Gmu-a0}. We remind the reader that the choice $\bar \alpha=\alphaz$ for $\Sigma^{}_{\NLO_{i+1}}\propto \alphas^n \bar{\alpha}(\alphaGmu^{m-n_\gamma}\alphaz^{n_\gamma})$ is in general inferior to $\bar \alpha=\alphaGmu$ and especially possible only if $n=0$, as shown in \eqref{eq:NLO_rescale_fixed}. The procedure for automating the implementation of mixed renormalisation according to  \eqref{eq:NLO_rescale_fixed} is actually the same than in the case  \eqref{eq:NLO_rescale}, but limited to $\bar \alpha=\alphaGmu$ and the usage of only the model {\tt loop\_qcd\_qed\_sm\_Gmu-a0}.

At this point,  it is worth to briefly remind the relations among the different renormalisation conditions. If we consider the $\alphaz$,   the $\alpha(\mz)$ and the $G_\mu$ schemes, these are the relations:
\begin{eqnarray}
\delta Z_e|_{G_\mu}&=& \delta Z_e|_{\alpha(0)}-\frac{1}{2}\Delta r\, ,\\
\delta Z_e|_{G_\mu}&=&\delta Z_e|_{\alpha(m_Z)}-\frac{1}{2}\left(-  \frac{c_W^2}{ s_W^2}\Delta \rho + \Delta r_{\rm rem}\right)\, ,\\
\delta Z_e|_{\alpha(\mz)}&=& \delta Z_e|_{\alpha(0)}-\frac{1}{2}\Delta\alpha(\mz^2)\,,
\label{alphaGmucond}
\end{eqnarray}
which obviously imply $\Delta r=\Delta\alpha(\mz^2)-\frac{c_W^2}{ s_W^2}\Delta \rho + \Delta r_{\rm rem}$ \cite{Sirlin:1980nh, Marciano:1980pb, Sirlin:1981yz}. The quantity $\Delta\alpha(\mz^2)$ is of purely QED origin and it takes into account the contribution of light fermions to the run of $\alpha$ from the scale $Q=0$ to $Q=\mz$, namely, 
\begin{equation}
\Delta\alpha(\mz^2)=\Pi^{AA}_{f\neq t}(0)-\Re\{\Pi^{AA}_{f\neq t}(\mz^2)\}.
\end{equation}
When fermions are treated as massless, $\Delta\alpha(\mz^2)$ exactly cancels the IR divergence in $\delta Z_e|_{\alpha(0)}$. The remaining components of $\Delta r$ are not IR sensitive and mainly concern the purely weak part of the renormalisation of $\alpha$. In particular, $\frac{c_W^2}{ s_W^2}\Delta \rho$ corresponds to the top-mass-enhanced corrections to the $\rho$ parameter.  After having recalled these differences we want to mention a possible discrepancy that may be left between a calculation in the $G_\mu$-scheme together with the photon fragmentation function and  the mixed  scheme with $\alphaz$ and $G_\mu$. While the running of the fragmentation function can naturally compensate for the effect of $\Delta\alpha(\mz^2)$,  the term $\Delta r - \Delta\alpha(\mz^2)$ has to be removed ``by hand'' in order to avoid its contribution to vertices with external photons.   

\subsubsection{Modification of the FKS counterterms}
\label{sec:fks}

Before discussing  the technical details concerning the IR counterterms for the integration of the separately divergent contributions of virtual and real emission diagrams, we remind the reader that
the  {\sc\small MadGraph5\_aMC@NLO} framework \cite{Alwall:2014hca} deals with IR singularities via the FKS method~\cite{Frixione:1995ms,
Frixione:1997np}, which has been automated for the first time in \mf~\cite{Frederix:2009yq,
Frederix:2016rdc}. We also recall that one-loop amplitudes can be evaluated via 
different types of integral-reduction techniques, the  OPP method~\cite{Ossola:2006us} or
 the Laurent-series expansion~\cite{Mastrolia:2012bu},
and  techniques for tensor-integral reduction~\cite{Passarino:1978jh,Davydychev:1991va,Denner:2005nn}.
All these  techniques are automated in the module \ml~\cite{Hirschi:2011pa}, which on top of  generating  the amplitudes switches dynamically among them. The codes \ct~\cite{Ossola:2007ax}, \nin~\cite{Peraro:2014cba,
Hirschi:2016mdz} and \coll~\cite{Denner:2016kdg} are employed within \ml, which has been optimised by taking inspiration from {\sc\small OpenLoops} \cite{Cascioli:2011va} for the integrand evaluation.

\medskip

We can now discuss the aforementioned IR counterterms. Since IR-divergent $\gamma\TO f \bar f$ splittings for isolated photons are vetoed and the IR structure of the renormalised one-loop amplitudes is altered when using the $\alphaz$-scheme, both the counterterms for regularising virtual and real contributions have to be altered. Regarding virtual contributions, in the FKS language this means modifying  the term $d\sigma^{(C,n)}$ (defined, {\it e.g.}, in eq.~3.26 of Ref.~\cite{Frederix:2018nkq}), which collects the Born-like remainders 
of the final- and initial-state collinear subtractions.\footnote{In this context $n$ is not the power of $\alphas$, but the number of final-state particles at Born.} Part of the modifications are due to the finite part ${\mathcal V}^{(n,1)}_{\rm FIN}$ of the virtual contribution, which in turn depends on what is included in the divergent part ${\mathcal V}^{(n,1)}_{\rm DIV}$ of the one-loop matrix
elements. As we have already discussed in details, by employing the $\alphaz$-scheme the IR-pole structure is altered w.r.t.~an $\MSbar$-like scheme. Therefore also ${\mathcal V}^{(n,1)}_{\rm DIV}$ has to be modified. Regarding the real radiation,  on the other hand,  nothing needs to be modified. Indeed, thanks to the implementation of the FKS subtraction method in  \mf~\cite{Frederix:2009yq}, by vetoing matrix elements stemming from the QED splitting of isolated photons, the corresponding real emission counterterm is not generated.

The quantity $d\sigma^{(C,n)}$ is defined at eqs.~(3.26--3.27) of Ref.~\cite{Frederix:2018nkq}, where NLO EW corrections and more in general Complete-NLO predictions have been automated, while ${\mathcal V}^{(n,1)}_{\rm DIV}$ at eqs.~(3.30--3.32) of the same reference. If ${\mathcal I}_k$  is an {\it isolated} photon (${\mathcal I}_k=\gamma_{\rm iso}$) then in the aforementioned equations 
\begin{equation}
C_T(\gamma_{\rm iso})=\gamma_T(\gamma_{\rm iso})=\gamma'_T(\gamma_{\rm iso})=0\,,\quad T={\rm QCD~or~QED}.
\end{equation}
While the condition $C_{\rm QED}(\gamma_{\rm iso})=0$ is unchanged w.r.t. ordinary photons, for which $C_{\rm QED}(\gamma)=0$ already holds, it is especially important to note that
\begin{eqnarray}
\gamma_{\rm QED}(\gamma_{\rm iso})&\ne&\gamma_{\rm QED}(\gamma)\,,\\
\gamma'_{\rm QED}(\gamma_{\rm iso})&\ne&\gamma'_{\rm QED}(\gamma)\,,
\end{eqnarray}
(the quantities on the r.h.s are defined in the appendix~A of Ref.~\cite{Frederix:2018nkq}).

We can now address a point that has been ignored so far in our discussion. Not all the vertices connected to photons in the final state have to be renormalised in the $\alphaz$-scheme. Indeed, this has to be done only if the photon is considered as a physical object, namely an isolated photon.
If for example one considers the process \eqref{eq:genproc} with $X$ containing jets, those have to be in general defined as democratic-jets and therefore photons can be part of them. At $\LO$, this means that each of those jets can be in principle formed by a single photon in the final state. It is very important to note that such photons are not isolated photons; they can split into fermions and especially their interactions with the rest of the process are renormalised in the $G_\mu$-scheme. However, for hadronic collisions,  the presence of final-state photons that can be tagged as a democratic jet  is very uncommon at $\LO_i$ for the case $i=1$. Indeed, since non-isolated photons and gluons are treated in the same way by the democratic-jet clustering, given a partonic process with a non-isolated photon in the final state a similar one with such a photon replaced by a gluon almost always exists.\footnote{This is not possible only if the process does not contain coloured particles in the final state and cannot be initiated by coloured partons.} Therefore, if the latter appears at $\LO_i$ for the final-state signature that is considered, the former appears at $\LO_{i+1}$.  On the other hand, this also means that in hadronic processes non-isolated photons can be in principle present at $\LO_i$ with $i>1$. Especially,  albeit being not very frequently, both isolated and non-isolated photons can be present at $\LO_i$ with $i>1$,  for instance in signatures featuring both isolated photons and jets. We leave this case for future work.

\subsubsection{Simultaneous photon isolation and democratic-jet clustering}
\label{sec:iso_clust}

While the simultaneous presence of both isolated and non-isolated photons is very uncommon at $\LO_1$ and not so frequent at $\LO_i$ with $i>1$, if isolated photons are present at $\LO_i$, both  isolated and non-isolated photons are always present at the same time at $\NLO_{i+1}$. Indeed, as soon as one external line or propagator in the process is electrically charged, the real emission of QED includes the process
\begin{equation}
p p \TO n_{\gamma} \gamma_{\rm iso} + X+\gamma\, . \label{eq:genprocrad}
\end{equation}  
This also means that, if light particles are part of $X$, democratic jets have to be  in general employed in order to achieve IR safety. If there are only leptons among the light particles of $X$, dressed leptons may be sufficient, but in general the main point is that non-isolated photons have to be recombined with massless particles when they get close to be collinear.

When both democratic jets and isolated photons are the physical objects appearing in the final state, the two algorithmic procedures for identifying them, isolation and clustering, do not commute.
If $X$ contains $n_j$ jets, the procedure that has to be followed for the inclusive cross section at $\NLO_{i}$ with $i>2$ for a process as defined as in \eqref{eq:genproc}, and therefore including also real radiation process as defined as in \eqref{eq:genprocrad}, is the following:
\begin{enumerate}
\item Run the photon-isolation algorithm, isolating photons from QCD-interacting particles as well as QED interacting particles, including photons themselves\footnote{For IR safety, the isolation of photons from photons is actually necessary only at NNLO and beyond.}.
\item If at least $n_\gamma$ photons are identified as isolated photons proceed, otherwise the event is rejected.
\item Run the jet clustering algorithm including all the QCD and QED interacting particles, but among the photons only those that have {\it not} been tagged as isolated.
\item If less than $n_j$ jets have been tagged, reject the event.

\end{enumerate}

If dressed leptons are part of the final-state physical objects, the recombination of bare leptons and non-isolated photons is done at the third step of the previous list. If both jet clustering and lepton recombination is performed, and the jet clustering involves non-isolated photons,  leptons and jets have to be  separated, {\it e.g.}, in the $(\eta,\phi)$ plane of the pseudorapidities and azimuthal angles.

 \section{Phenomenological results for top-quark and photon associated production modes}
 \label{sec:processes}
 
 \subsection{Common set-up}
 
 \label{sec:setup}

In this section we describe the calculation setup, which is common for the processes we have considered in this work:
\begin{itemize}
\item $pp~\TO ~t\bar t \gamma$,
\item $pp~\TO ~t\bar t \gamma \gamma $,
\item $pp~\TO~ t\gamma j+\bar t\gamma j$,
\item $\tadlep  $ and $\tadhad$.
\end{itemize} 
Unless it is differently specified, in the following with the notation $t\gamma j$ we will understand both $t\gamma j$ and $\bar t\gamma j$ production. Also, we will understand that $\gamma$ is an isolated photon, without specifying $\gamma_{\rm iso}$ as in the previous sections.  
We provide results for proton--proton collisions at the LHC, with a centre-of-mass energy of 13 TeV.
In our calculation, we employ the complex mass scheme \cite{Denner:1999gp,Denner:2005fg,Frederix:2018nkq}, using the following on-shell input parameters
\begin{align}
m_{Z}&=91.188~\textrm{GeV}\, ,&\quad  m_{W}&=80.385~\textrm{GeV}\, ,&\quad  m_H&=125~\textrm{GeV}\, ,& \nonumber \\
   m_{\textrm{t}}&=173.3~\textrm{GeV}\, ,&\quad m_{b}&=4.92~\textrm{GeV}\, ,&\quad \Gamma_{\textrm{t}}&=0\, , &\\ 
\Gamma_{Z}&=2.49707~\textrm{GeV}\, ,&\quad \Gamma_{W}&=2.09026~\textrm{GeV}\, ,& \Gamma_H &=4.07902~\textrm{GeV}\, .&\nonumber
\end{align}
We have set $\Gamma_{t}=0$, since at least one external top quark is always present. In the case of $\tta$ and $\ttaa$ production, also the widths of the $W$ and $Z$ bosons are set equal to zero.
All the calculations are performed in the five-flavour scheme (5FS), besides the case of $\taj$ production, where also the NLO QCD calculation in the four-flavour scheme (4FS) is considered for estimating the flavour-scheme uncertainty.  The value $m_{b}=4.92~\textrm{GeV}$ directly enters  the calculation only in this specific case and has been chosen in order to be consistent with the corresponding calculation in the 5FS.  Following the same argument of Ref.~\cite{Pagani:2020mov}, we choose the set {\sc\small NNPDF3.1} \cite{Ball:2017nwa, Bertone:2017bme}  for all our calculations. In this set, the value of $m_{b}$ used in the PDF evolution is precisely $m_{b}=4.92~\textrm{GeV}$.
   
As we have discussed in Sec.~\ref{sec:implementation},  we  renormalised  EW interactions in a mixed scheme. The input values for $G_\mu$ and $\alphaz$ are: 
\begin{equation}
    \gmu = 1.16639 \cdot 10^{-5} ~\gev^{-2}\,,\qquad \alphaz=\frac{1}{137.036}\,.
\end{equation}
QCD interactions are instead renormalised  in the $\MSbar$-scheme, with  the (renormalisation-group running) value of $\alphas$ directly taken from the PDF sets used in the calculation. We estimate QCD scale uncertainties by  independently varying  by a factor of two both the renormalisation scale $\mu_r$ and the factorisation scale $\mu_f$ around the central value $\mu_0$ defined as follows,
\begin{align}
 \mu_0\equiv H_T/6&=\frac{\sum_i m_{T,i}}{6}\,,~~~  i=t,\gamma,j_b&{\rm for~}\taj \, ,   \label{eq:scaletaj}    \\ 
 \mu_0\equiv H_T/2&=\frac{\sum_i m_{T,i}}{2}&{\rm for~the~other~production~processes} \, .   \label{eq:scaletaj2}    
\end{align}
The quantity $m_{T,i}$ is the transverse mass of the particle $i$.
 The scale definition in eq.~\eqref{eq:scaletaj}, where with $j_b$ we denote the $b$-jet, is analogue to the one of Ref.~\cite{Pagani:2020mov}, which is based on the findings of Refs.~\cite{Maltoni:2012pa, Demartin:2015uha}. The definition  in eq.~\eqref{eq:scaletaj2} is instead the default option in \mglong, with the sum running over the final-state-particle momenta, including those from real emissions.

Finally, we specify the parameters related to procedure explained in Sec.~\ref{sec:iso_clust} for the isolation of photons and the clustering of democratic jets or dressed leptons.
Photon isolation is performed {\it  \`a la} Frixione \cite{Frixione:1998jh}, with the parameters
\begin{equation}
R_0(\gamma)=0.4\,,\qquad\epsilon_\gamma=1\,,\qquad n=1\, ,\qquad p^{\rm min}_T(\gamma)>25~\gev\,. \label{eq:FP}
\end{equation}

 After this, we cluster jets via the anti-$k_T$ algorithm \cite{Cacciari:2008gp}  as implemented in {\sc \small FastJet}~\cite{Cacciari:2011ma} using the parameters 
\begin{equation}
p_T^{\rm min}=40~{\rm GeV}\,, \qquad R=0.4\,. 
\end{equation}
We remind the reader that in our calculation a jet can correspond to a single non-isolated photon.\footnote{LHC analyses typically defines jets with up to 99\% of their energy of electromagnetic origin. Up to 90\%  can even be associated to a single photon. See also  Ref.~\cite{Frederix:2016ost}.} 
 When we will consider $b$-jets, in the case of $\taj$ production, we will simply mean  jets containing a bottom (anti)quark; no restrictions on their pseudorapidity are imposed.\footnote{In our calculation, no $\gamma,g\TO b \bar b$ splittings are involved in the final state and in turn, $b$-jets cannot include more than one bottom (anti)quark. Therefore, no IR safety problems are present in this $b$-jet definition even if we use the  5FS.} Also, for this process, the jet definition is relevant only for differential distributions and not for total cross sections; single-top photon is properly defined and IR finite without tagging any jet.

 In Sec.~\ref{sec:decay} we will also deal with leptons in the final state, which have to be dressed with photons in order to achieve IR safety. Since in this work lepton--photon recombination concerns only the case of top-quark decays in their rest frame, a dressed lepton is obtained by recombining a bare lepton $\ell$ with any non-isolated photon $\gamma$ satisfying the condition 
\begin{equation}
\Delta \theta(\ell, \gamma) < 0.05\,, \label{eq:recQED}
\end{equation}
 where  $\Delta \theta(\ell, \gamma)$ is the angle between the lepton and the photon.  For a general production process this procedure can be reframed via $\Delta R(\ell, \gamma)$ in place of $\Delta \theta(\ell, \gamma)$, where   
 $\Delta R(\ell, \gamma) \equiv \sqrt{(\Delta \eta(\ell, \gamma))^2+(\Delta \phi(\ell, \gamma))^2} $ and  $\Delta \eta(\ell, \gamma)$ and $\Delta \phi(\ell, \gamma)$ are the difference of the bare-lepton and photon pseudorapidities and azimuthal angles, respectively.  In case  the recombination condition is satisfied for more than one bare lepton, the photon is clustered together with the one for which $\Delta R(\ell, \gamma)$ is the smallest.

\subsection{Top-quark pair and one photon associated production: $\tta$ }
\label{sec:tta}

The NLO EW corrections to top-quark pair hadroproduction in association with a single photon ($\tta$) have already been calculated in Ref.~\cite{Duan:2016qlc}, by using the $\alphaz$-scheme. We repeat the calculation, in a completely automated way, by employing the mixed renormalisation scheme discussed in Sec.~\ref{sec:set-up} and providing for the first time Complete-NLO predictions. For this process, according to eq.~\eqref{eq:blobs_NLO_general}, $k=3$ and therefore not only NLO EW and NLO QCD corrections are present ($\NLO_1$ and $\NLO_2$ in our notation), but also the $\LO_2$, $\LO_3$, $\NLO_3$ and $\NLO_4$ contributions,  where the $\LO_1$ is proportional to $\alpha_s^2\alpha$.

We remind the reader that NLO QCD corrections to $\tta$ production have been already calculated in Refs.~\cite{PengFei:2009ph,PengFei:2011qg,Maltoni:2015ena, Mangano:2016jyj}, and in particular in Refs.~\cite{Maltoni:2015ena, Mangano:2016jyj} it has been shown their large impact in reducing the top-quark charge asymmetry at the LHC. This last aspect has also been investigated in Ref.~\cite{Bergner:2018lgm}. The matching with QCD parton shower, besides being in general available in the {\mglong} framework and taken into account in Ref.~\cite{Maltoni:2015ena}, has been studied  in Ref.~\cite{Kardos:2014zba}, without spin correlations, via the \textsc{PowHel}
framework \cite{Bevilacqua:2011xh}, which in turn relies on the
\textsc{Powheg-Box} system \cite{Frixione:2007vw, Alioli:2010xd}. NLO QCD corrections including top-quark decays have been presented for the first time in Ref.~\cite{Melnikov:2011ta} in the narrow width approximation (NWA), and for the complete non-resonant $e^+\nu_e\mu^-\nu_\mu
b\bar{b}\gamma$ leptonic signature in Ref.~\cite{Bevilacqua:2018woc}. Comparison among the different NLO QCD approximations has been carried out in Ref.~\cite{Bevilacqua:2019quz}.

This process has already been observed at the LHC \cite{Aad:2015uwa}, and further measurements have been performed \cite{Aaboud:2017era, Sirunyan:2017iyh, Aaboud:2018hip, Aad:2020axn}, showing so far no sign of deviations from the SM predictions.

\subsubsection{Numerical results }

\begin{table}[!t]
\renewcommand{\arraystretch}{2.5}
\scriptsize
\begin{center}
\begin{tabular}{c c | c c }
\hline
\hline
\multicolumn{4}{c}{{\normalsize $t \bar t \gamma$ }} \\
\hline
Cuts & Order & $\sigma$ [fb] & $A_C$ [\%]\\
\hline
\multirow{4}{3.1cm}{$p_T(\gamma) \ge 25$~GeV}  & LO$_{\rm QCD}$ & $1100(1)_{-232.13 (-21.1  \%)}^{+321.82 (+29.3  \%)}~_{-12.02 (-1.1  \%)}^{+12.02 (+1.1  \%)}$ & $-4.14(8)_{-0.19 (-4.7  \%)}^{+0.21 (+5.0  \%)}~_{-0.14 (-3.3  \%)}^{+0.14 (+3.3  \%)}$ \\
  & NLO$_{\rm QCD}$ & $1743(6)_{-214.16 (-12.3  \%)}^{+215.41 (+12.4  \%)}~_{-15.96 (-0.9  \%)}^{+15.96 (+0.9  \%)}$ & $-2.1(1)_{-0.35 (-16.6  \%)}^{+0.45 (+21.3  \%)}~_{-0.11 (-5.1  \%)}^{+0.11 (+5.1  \%)}$ \\
  & NLO$_{\rm QCD+EW}$ & $1720(6)_{-207.97 (-12.1  \%)}^{+206.53 (+12.0  \%)}~_{-15.87 (-0.9  \%)}^{+15.87 (+0.9  \%)}$  & $-1.9(1)_{-0.38 (-19.8  \%)}^{+0.49 (+25.7  \%)}~_{-0.12 (-6.5  \%)}^{+0.12 (+6.5  \%)}$ \\
  & NLO & $1744(6)_{-209.78 (-12.0  \%)}^{+209.19 (+12.0  \%)}~_{-17.46 (-1.0  \%)}^{+17.46 (+1.0  \%)}$  & $-1.8(1)_{-0.36 (-20.1  \%)}^{+0.48 (+26.4  \%)}~_{-0.13 (-7.1  \%)}^{+0.13 (+7.1  \%)}$ \\
\hline
\multirow{4}{3.1cm}{$p_T(\gamma) \ge 50$~GeV}  & LO$_{\rm QCD}$ & $574.5(4)_{-123.76 (-21.5  \%)}^{+172.60 (+30.0  \%)}~_{-6.28 (-1.1  \%)}^{+6.28 (+1.1  \%)}$ & $-4.00(7)_{-0.19 (-4.8  \%)}^{+0.20 (+5.1  \%)}~_{-0.13 (-3.3  \%)}^{+0.13 (+3.3  \%)}$  \\
  & NLO$_{\rm QCD}$ & $912(5)_{-113.94 (-12.5  \%)}^{+113.87 (+12.5  \%)}~_{-9.17 (-1.0  \%)}^{+9.17 (+1.0  \%)}$ & $-2.2(1)_{-0.33 (-15.2  \%)}^{+0.42 (+19.1  \%)}~_{-0.11 (-4.8  \%)}^{+0.11 (+4.8  \%)}$ \\
  & NLO$_{\rm QCD+EW}$ & $900(5)_{-110.58 (-12.3  \%)}^{+109.04 (+12.1  \%)}~_{-8.93 (-1.0  \%)}^{+8.93 (+1.0  \%)}$ & $-2.0(1)_{-0.35 (-17.7  \%)}^{+0.45 (+22.6  \%)}~_{-0.12 (-6.1  \%)}^{+0.12 (+6.1  \%)}$ \\
  & NLO & $912(5)_{-111.56 (-12.2  \%)}^{+110.44 (+12.1  \%)}~_{-10.00 (-1.1  \%)}^{+10.00 (+1.1  \%)}$ & $-1.9(1)_{-0.34 (-17.9  \%)}^{+0.44 (+23.1  \%)}~_{-0.13 (-6.8  \%)}^{+0.13 (+6.8  \%)}$ \\
\hline
\multirow{4}{3.1cm}{$p_T(\gamma) \ge 25$~GeV, $|y(\gamma)| \le 2.5$}  & LO$_{\rm QCD}$ & $1025(1)_{-216.96 (-21.2  \%)}^{+301.02 (+29.4  \%)}~_{-10.56 (-1.0  \%)}^{+10.56 (+1.0  \%)}$ & $-4.00(8)_{-0.20 (-4.9  \%)}^{+0.21 (+5.2  \%)}~_{-0.13 (-3.2  \%)}^{+0.13 (+3.2  \%)}$ \\
  & NLO$_{\rm QCD}$ & $1559(2)_{-181.06 (-11.6  \%)}^{+171.64 (+11.0  \%)}~_{-15.31 (-1.0  \%)}^{+15.31 (+1.0  \%)}$ & $-2.1(1)_{-0.32 (-14.8  \%)}^{+0.40 (+18.8  \%)}~_{-0.10 (-4.7  \%)}^{+0.10 (+4.7  \%)}$ \\
  & NLO$_{\rm QCD+EW}$ & $1537(2)_{-175.13 (-11.4  \%)}^{+163.15 (+10.6  \%)}~_{-14.90 (-1.0  \%)}^{+14.90 (+1.0  \%)}$ & $-1.9(1)_{-0.34 (-17.4  \%)}^{+0.43 (+22.3  \%)}~_{-0.12 (-6.1  \%)}^{+0.12 (+6.1  \%)}$ \\
  & NLO & $1557(2)_{-176.51 (-11.3  \%)}^{+165.26 (+10.6  \%)}~_{-16.45 (-1.1  \%)}^{+16.45 (+1.1  \%)}$ & $-1.9(1)_{-0.33 (-17.5  \%)}^{+0.42 (+22.8  \%)}~_{-0.13 (-6.7  \%)}^{+0.13 (+6.7  \%)}$ \\
\hline
\multirow{4}{3.1cm}{$p_T(\gamma) \ge 50$~GeV, $|y(\gamma)| \le 2.5$}  & LO$_{\rm QCD}$ & $547.1(4)_{-118.00 (-21.6  \%)}^{+164.60 (+30.1  \%)}~_{-6.17 (-1.1  \%)}^{+6.17 (+1.1  \%)}$ & $-3.87(7)_{-0.19 (-5.0  \%)}^{+0.20 (+5.3  \%)}~_{-0.13 (-3.4  \%)}^{+0.13 (+3.4  \%)}$ \\
  & NLO$_{\rm QCD}$ & $843(1)_{-101.09 (-12.0  \%)}^{+96.77 (+11.5  \%)}~_{-8.51 (-1.0  \%)}^{+8.51 (+1.0  \%)}$ & $-2.2(1)_{-0.32 (-14.4  \%)}^{+0.40 (+18.1  \%)}~_{-0.10 (-4.6  \%)}^{+0.10 (+4.6  \%)}$ \\
  & NLO$_{\rm QCD+EW}$ & $831(1)_{-97.83 (-11.8  \%)}^{+92.10 (+11.1  \%)}~_{-8.29 (-1.0  \%)}^{+8.29 (+1.0  \%)}$ & $-2.0(1)_{-0.34 (-16.5  \%)}^{+0.43 (+21.1  \%)}~_{-0.11 (-5.6  \%)}^{+0.11 (+5.6  \%)}$ \\
  & NLO & $842(1)_{-98.62 (-11.7  \%)}^{+93.26 (+11.1  \%)}~_{-9.50 (-1.1  \%)}^{+9.50 (+1.1  \%)}$ & $-2.0(1)_{-0.33 (-16.7  \%)}^{+0.42 (+21.5  \%)}~_{-0.12 (-6.3  \%)}^{+0.12 (+6.3  \%)}$ \\
\hline
\hline
\end{tabular}
\end{center}
\caption{Cross sections and charge asymmetries for $t \bar t \gamma$ production.  The uncertainties are respectively the scale and the PDF ones in the form: $\pm$ absolute size ($\pm$ relative size). The first number in parentheses after the central value is the absolute statistical error. \label{tab:tta}}  
\end{table}

In Tab.~\ref{tab:tta} we provide results for the total cross section and the charge asymmetry $A_C$, with different cuts on the transverse momentum and rapidity of the photon. We remind the reader that the charge asymmetry is defined as 
 \begin{equation}
A_C=\frac{\sigma(|y_t|>|y_{\bar{t}}|)-\sigma(|y_t|<|y_{\bar{t}}|)}{\sigma(|y_t|>|y_{\bar{t}}|)+\sigma(|y_t|<|y_{\bar{t}}|)}\,.\label{eq:asym}
\end{equation}
In all cases results are provided in different approximations, namely, 
\begin{eqnarray}
\LOQCD&\equiv&\LO_1\, , \label{eq:alias1}\\
\NLOQCD&\equiv&\LO_1 + \NLO_1 \, , \\
\NLOQCDEW&\equiv&\LO_1 + \NLO_1+ \NLO_2 \, ,\\
\NLO&\equiv& \LO_1 + \LO_2 + \LO_3 + \label{eq:alias4}\\
&&\NLO_1 + \NLO_2 + \NLO_3 +  \NLO_4 \, ,\nonumber
\end{eqnarray}
where in  eqs.~\eqref{eq:alias1}--\eqref{eq:alias4} there are the precise definitions of the quantities entering tables and plots in this section. The Complete-NLO is therefore simply denoted as ``NLO''. 
Moreover, for the case with the cut $p_T(\gamma) \ge 25$ GeV, we show in Tab.~\ref{tab:tta_ratio} the ratio of the contribution of each separate perturbative order with the $\LOQCD$.

As can be seen in Tab.~\ref{tab:tta}, for all the four phase-space cuts choices, NLO EW corrections are negative and $\sim -2 \%$ of the $\LOQCD$ or equivalently $\sim -1 \%$ of the $\NLOQCD$, {\it i.e.}, one order of magnitude smaller than QCD scale uncertainties at NLO accuracy. Moreover, the difference between the Complete-NLO prediction, NLO in the table, and the $\NLOQCD$ one is even smaller. Indeed, as can be seen in  Tab.~\ref{tab:tta_ratio} for the case $p_T(\gamma) \ge 25$ GeV,  the $\LO_2$, the $\LO_3$ and the $\NLO_3$ all together largely cancel the impact of the $\NLO_2$, the NLO EW corrections. Our conclusion is that,  given the current QCD uncertainties (scale+PDF),  which are dominated by the scale dependence at NLO, {\it at the inclusive level} the impact of EW corrections on the $\tta$ cross section is negligible. We remind the reader that this conclusion could be drawn only after having performed a complete calculation. Moreover, it is in contrast to what has been observed for other processes involving top quarks, such as $t \bar t W$ and $t \bar t t \bar t$ production \cite{Frederix:2017wme}. 

\begin{table}[!t]
\renewcommand{\arraystretch}{2.25}
\begin{center}
\begin{tabular}{c| c  c c c c c c }
\hline
\hline
\multicolumn{7}{c}{$t \bar t \gamma$  ($p_T(\gamma) \ge 25$ GeV) } \\
\hline
Order                     & $\LO_2$       & $\LO_3$      & $\NLO_1$ & $\NLO_2$ & $\NLO_3$ & $\NLO_4$  \\
\hline
ratio over $\LO_1$ [\%] &  0.2 & 1.1 & 58.6 & -2.1 & 0.8 & < 0.1  \\
\hline
\hline
\end{tabular}
\end{center}
\caption{Relative contribution of perturbative orders entering Complete-NLO predictions for $t \bar t \gamma$ production with $p_T(\gamma) \ge 25$ GeV.\label{tab:tta_ratio}}  
\end{table}

The impact of the NLO corrections is different in the case of the charge asymmetry $A_C$. First of all, the NLO QCD corrections strongly decrease the $\LOQCD$ predictions, as already discussed in Refs.~\cite{Maltoni:2015ena,Bergner:2018lgm}, with the  $\NLOQCD/\LOQCD$ ratio ranging from 0.51 to 0.57 in the four phase-space cuts choices of Tab.~\ref{tab:tta}. Moreover,  since we evaluate scale uncertainties by keeping scales correlated in the numerator and denominator of $A_C$ (see  eq.~\eqref{eq:asym}),  $\LOQCD$ scale uncertainties are very small. However, NLO QCD corrections induce an additional negative term to the numerator of $A_C$, which therefore has a scale dependence that is anti-correlated with the one of the denominator. The net effect is an increment, but also a more realistic estimate, of the scale uncertainty for $A_C$. The impact of NLO EW corrections is also not negligible, with the $\NLOQCDEW/\NLOQCD$ ratio ranging from 0.89 to 0.92  in the four phase-space cuts choices. The additional terms in the Complete-NLO,  $(\NLO-\NLOQCDEW)$, further reduce the predictions,  with the $\NLO/\NLOQCD$ ratio ranging from 0.84 to 0.88 in the four phase-space cuts choices. Overall, the Complete-NLO predictions reduce the $\NLOQCD$ ones by shifting their central values to (almost) the lower edge of the scale-uncertainty bands.

\begin{figure}[!t]
\centering
\includegraphics[width=0.325\textwidth]{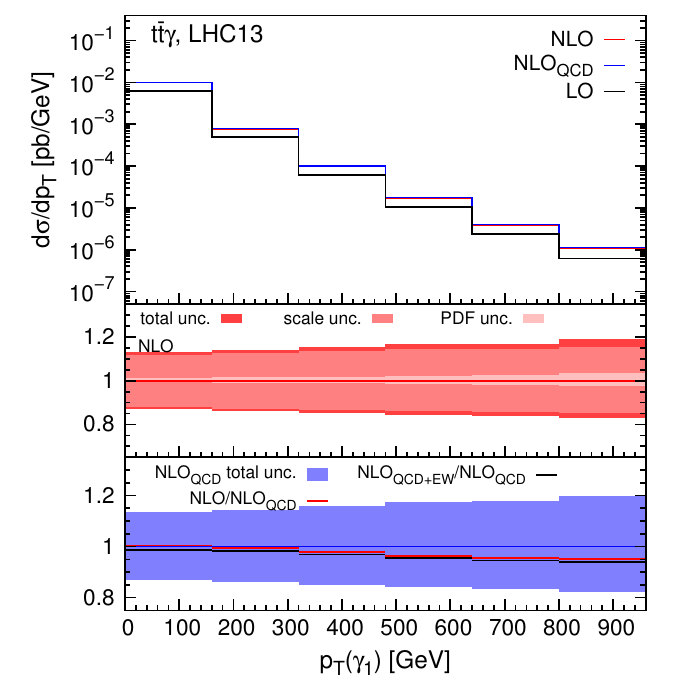}
\includegraphics[width=0.325\textwidth]{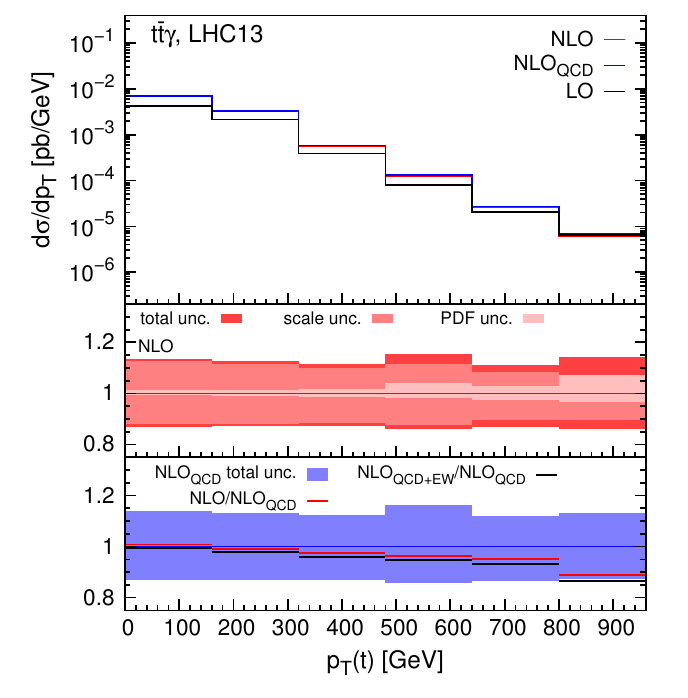}
\includegraphics[width=0.325\textwidth]{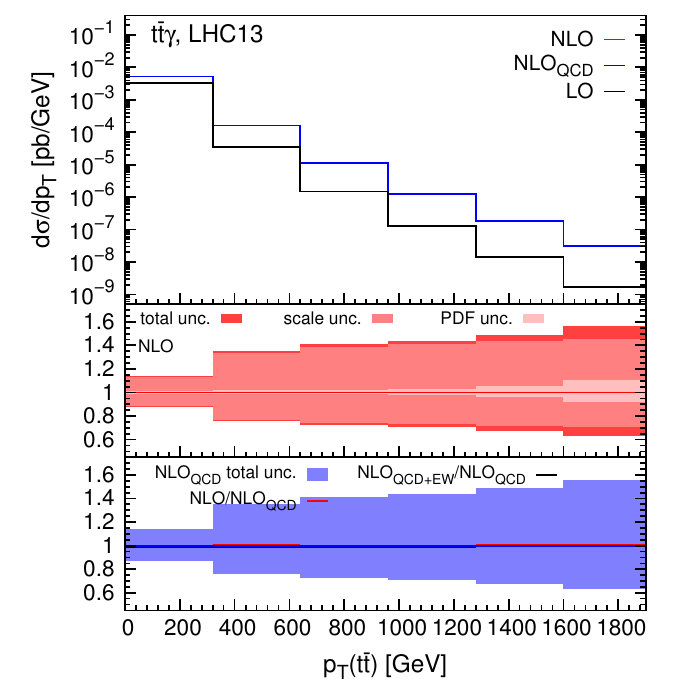}
\includegraphics[width=0.325\textwidth]{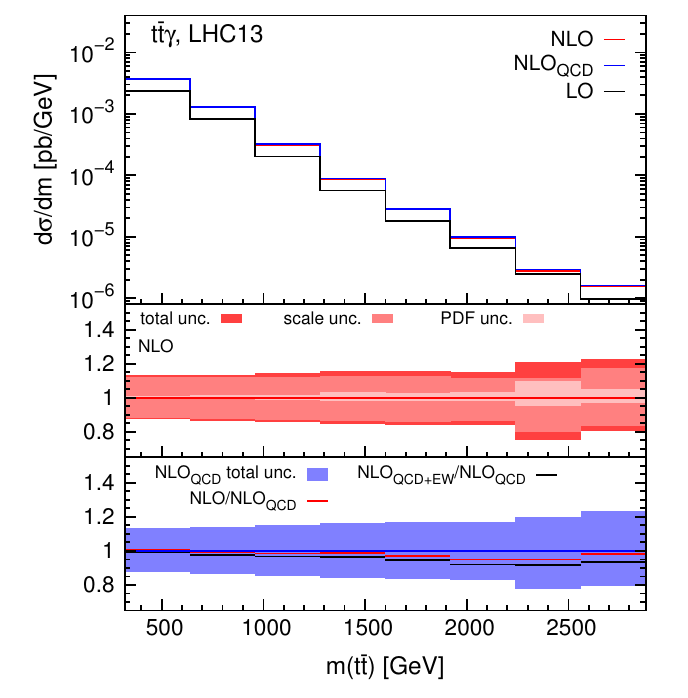}
\includegraphics[width=0.325\textwidth]{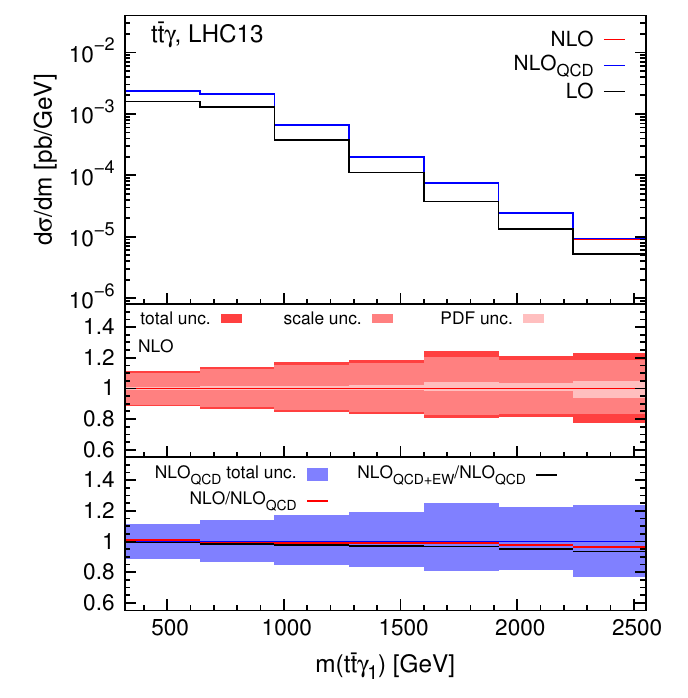}
\includegraphics[width=0.325\textwidth]{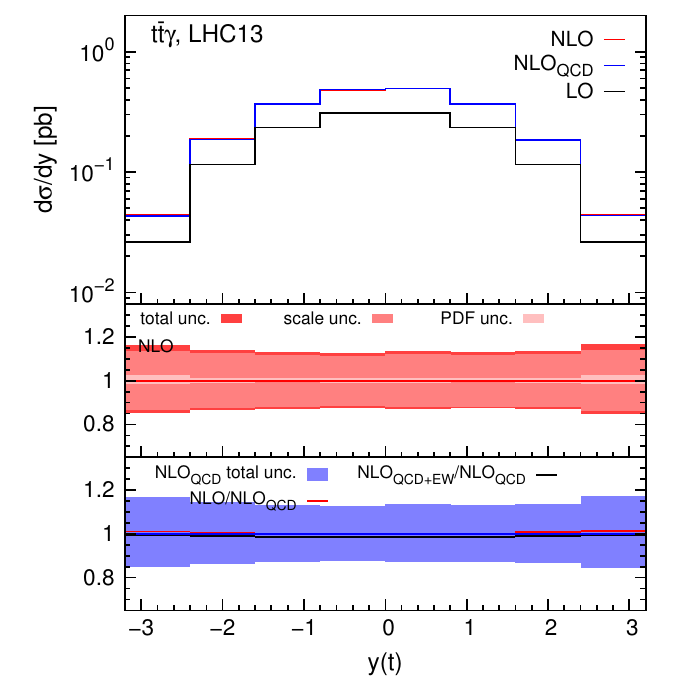}

\caption{Differential distributions for $\tta$ production. \label{fig:tta} }
\end{figure}

In Fig.~\ref{fig:tta} we show differential distributions for $\tta$ production. In particular we show the transverse momentum distributions ($p_T$) of the hardest isolated photon ($\gamma_1$),  the top quark and the top-quark pair, the invariant mass of the  top-quark pair and of the entire $\tta$ system, and the rapidity of the top-quark. For each plot we show in the main panel the central value of the LO, $\NLOQCD$ and $\NLO$ predictions. In the first inset we separately show the relative scale and PDF uncertainties of the NLO prediction together with their sum in quadrature, the total uncertainty. In the last inset we show again the relative total uncertainty, but now for the NLO QCD prediction, together with the $\NLOQCDEW/\NLOQCD$ and $\NLO/\NLOQCD$ ratios.

For the $p_T(\gamma)$ and $p_T(t)$ distributions, the NLO EW corrections are negative and grow in absolute value in the tail. This effect is expected and due to the EW Sudakov logarithms. For these two observables, the impact of the NLO EW corrections cannot be neglected, especially for $p_T(t)$, where in the tail the term $\NLO_2=(\NLOQCDEW-\NLOQCD)$  is almost as large as the total $\NLOQCD$ uncertainty, which in turn, as for any other observable considered here, is numerically as large as the total $\NLO$ uncertainty. We also notice that the impact of the $(\NLO-\NLOQCDEW)$ term is on the other hand negligible.
The case of $p_T(t \bar t)$ is special. As discussed in detail in Refs.~\cite{Maltoni:2015ena, Mangano:2016jyj} and also visible in the main panel of the $p_T(t \bar t)$ plot in Fig.~\ref{fig:tta}, the NLO QCD corrections scale as $\alphas\log^2(p_T(t \bar t )/Q)$ where $Q$ is a scale that increases by increasing $R_0(\gamma)$ or  $p^{\rm min}_T(\gamma)$, the isolation parameters of eq.~\eqref{eq:FP}. This effect underlies the large increase of scale and PDF uncertainties and the smallness of the $(\NLOQCDEW/\NLOQCD-1)$ and $(\NLO/\NLOQCD-1)$ terms.
For what concerns the $m(t \bar t)$ and $m(\tta)$ distributions, we see similar effects as in the  $p_T(\gamma)$ and $p_T(t)$ ones for the  $\NLOQCDEW/\NLOQCD$ ratio, although with smaller deviations from unity. On the other hand, especially for $m(t \bar t)$, the effect is largely compensated by the additional terms in $(\NLO-\NLOQCDEW)$. The $y(t)$ rapidity does not show large EW effects, similarly to the inclusive rates. The only effects that are not flat are the relative size of the uncertainties, growing in the peripheral region.

\subsection{Top-quark pair and two photons associated production: $\ttaa$ }

\begin{table}[!t]
\renewcommand{\arraystretch}{2.5}
\scriptsize
\begin{center}
\begin{tabular}{c c | c c }
\hline
\hline
\multicolumn{4}{c}{\normalsize $t \bar t \gamma \gamma$ } \\
\hline
Cuts & Order & $\sigma$ [fb] & $A_C$ [\%]\\
\hline
\multirow{3}{3.1cm}{$p_T(\gamma_{1,2}) \ge 25$ GeV, $\Delta R(\gamma_1,\gamma_2) \ge 0.4$} & LO$_{\rm QCD}$ & $3.20(1)_{-0.65 (-20.4  \%)}^{+0.90 (+28.1  \%)}~_{-0.05 (-1.4  \%)}^{+0.05 (+1.4  \%)}$ & $-18.8(2)_{-0.44 (-2.4  \%)}^{+0.50 (+2.7  \%)}~_{-0.58 (-3.1  \%)}^{+0.58 (+3.1  \%)}$ \\
  & NLO$_{\rm QCD}$ & $5.09(5)_{-0.63 (-12.5  \%)}^{+0.67 (+13.2  \%)}~_{-0.06 (-1.2  \%)}^{+0.06 (+1.2  \%)}$ & $-12(1)_{-1.31 (-10.6  \%)}^{+1.75 (+14.2  \%)}~_{-0.45 (-3.6  \%)}^{+0.45 (+3.6  \%)}$ \\
  & NLO$_{\rm QCD+EW}$ & $4.95(5)_{-0.60 (-12.1  \%)}^{+0.62 (+12.6  \%)}~_{-0.06 (-1.2  \%)}^{+0.06 (+1.2  \%)}$ & $-12(1)_{-1.39 (-11.9  \%)}^{+1.89 (+16.0  \%)}~_{-0.47 (-4.0  \%)}^{+0.47 (+4.0  \%)}$ \\
\hline
\multirow{3}{3.1cm}{$p_T(\gamma_{1,2}) \ge 50$~GeV, $\Delta R(\gamma_1,\gamma_2) \ge 0.4$}  & LO$_{\rm QCD}$ & $0.92(1)_{-0.19 (-21.1  \%)}^{+0.27 (+29.3  \%)}~_{-0.01 (-1.6  \%)}^{+0.01 (+1.6  \%)}$ & $-17.8(2)_{-0.48 (-2.7  \%)}^{+0.54 (+3.0  \%)}~_{-0.63 (-3.6  \%)}^{+0.63 (+3.6  \%)}$ \\
  & NLO$_{\rm QCD}$ & $1.47(1)_{-0.19 (-12.7  \%)}^{+0.19 (+13.2  \%)}~_{-0.02 (-1.3  \%)}^{+0.02 (+1.3  \%)}$ & $-11.5(7)_{-1.16 (-10.1  \%)}^{+1.50 (+13.0  \%)}~_{-0.49 (-4.2  \%)}^{+0.49 (+4.2  \%)}$ \\
  & NLO$_{\rm QCD+EW}$ & $1.43(1)_{-0.18 (-12.4  \%)}^{+0.18 (+12.6  \%)}~_{-0.02 (-1.4  \%)}^{+0.02 (+1.4  \%)}$ & $-11.0(7)_{-1.24 (-11.2  \%)}^{+1.62 (+14.6  \%)}~_{-0.54 (-4.9  \%)}^{+0.54 (+4.9  \%)}$ \\
\hline
\multirow{3}{3.1cm}{$p_T(\gamma_{1,2}) \ge 25$~GeV, $|y(\gamma_{1,2})| \le 2.5$, $\Delta R(\gamma_1,\gamma_2) \ge 0.4$}  & LO$_{\rm QCD}$ & $2.67(1)_{-0.55 (-20.5  \%)}^{+0.76 (+28.2  \%)}~_{-0.03 (-1.3  \%)}^{+0.03 (+1.3  \%)}$ & $-17.3(1)_{-0.47 (-2.7  \%)}^{+0.53 (+3.1  \%)}~_{-0.48 (-2.8  \%)}^{+0.48 (+2.8  \%)}$ \\
  & NLO$_{\rm QCD}$ & $4.04(3)_{-0.47 (-11.6  \%)}^{+0.46 (+11.4  \%)}~_{-0.05 (-1.2  \%)}^{+0.05 (+1.2  \%)}$ & $-13.1(8)_{-0.93 (-7.1  \%)}^{+1.17 (+9.0  \%)}~_{-0.37 (-2.8  \%)}^{+0.37 (+2.8  \%)}$ \\
  & NLO$_{\rm QCD+EW}$ & $3.91(3)_{-0.44 (-11.2  \%)}^{+0.42 (+10.7  \%)}~_{-0.05 (-1.2  \%)}^{+0.05 (+1.2  \%)}$ & $-12.7(8)_{-0.99 (-7.8  \%)}^{+1.27 (+10.0  \%)}~_{-0.39 (-3.1  \%)}^{+0.39 (+3.1  \%)}$ \\
\hline
\multirow{3}{3.1cm}{$p_T(\gamma_{1,2}) \ge 50$~GeV, $|y(\gamma_{1,2})| \le 2.5$, $\Delta R(\gamma_1,\gamma_2) \ge 0.4$}  & LO$_{\rm QCD}$ & $0.82(1)_{-0.17 (-21.1  \%)}^{+0.24 (+29.3  \%)}~_{-0.01 (-1.6  \%)}^{+0.01 (+1.6  \%)}$ & $-16.7(2)_{-0.49 (-3.0  \%)}^{+0.55 (+3.3  \%)}~_{-0.53 (-3.2  \%)}^{+0.53 (+3.2  \%)}$ \\
  & NLO$_{\rm QCD}$ & $1.28(1)_{-0.16 (-12.2  \%)}^{+0.16 (+12.3  \%)}~_{-0.02 (-1.3  \%)}^{+0.02 (+1.3  \%)}$ & $-10.1(6)_{-1.16 (-10.9  \%)}^{+1.51 (+14.2  \%)}~_{-0.42 (-3.9  \%)}^{+0.42 (+3.9  \%)}$ \\
  & NLO$_{\rm QCD+EW}$ & $1.24(1)_{-0.15 (-11.9  \%)}^{+0.15 (+11.7  \%)}~_{-0.02 (-1.3  \%)}^{+0.02 (+1.3  \%)}$ & $-10.2(7)_{-1.24 (-12.1  \%)}^{+1.63 (+15.9  \%)}~_{-0.43 (-4.3  \%)}^{+0.43 (+4.3  \%)}$ \\
\hline
\hline
\end{tabular}
\end{center}
\caption{Cross sections and charge asymmetries for $t \bar t \gamma \gamma$ production.  The uncertainties are respectively the scale and the PDF ones in the form: $\pm$ absolute size ($\pm$ relative size). The first number in parentheses after the central value is the absolute statistical error.} 
\label{tab:ttaa} 
\end{table}

The calculation of NLO EW corrections to top-quark pair hadroproduction in association with two  photons ($\ttaa$) is presented for the first time here. We perform the calculation, in a completely automated way, and we limit ourselves to the case of NLO EW and NLO QCD corrections. However, also for this  process, according to eq.~\eqref{eq:blobs_NLO_general}, $k=3$ and therefore not only NLO EW and NLO QCD corrections are present ($\NLO_1$ and $\NLO_2$ in our notation). We leave the Complete-NLO study to future work, but given  what has been observed in the case of $\tta$ production, we do not expect large effects in comparison to the QCD uncertainties. 

We remind the readers that NLO QCD corrections to $\ttaa$ production have been calculated for the first time in Ref.~\cite{Kardos:2014pba}, matched to parton shower effects in Ref.~\cite{vanDeurzen:2015cga} and thoroughly studied together with all the other $t \bar t VV$ processes in Ref.~\cite{Maltoni:2015ena}. The last two references have also investigated its impact in the $t \bar t H$ searches where the Higgs boson decays into two photons, which is one of the main motivations to study  $\ttaa$ production at the LHC. 

\subsubsection{Numerical results}

Similarly to the case of $\tta$ in Tab.~\ref{tab:tta}, in Tab.~\ref{tab:ttaa} we provide results for the total cross section and the charge asymmetry $A_C$ for $\ttaa$ production, with different cuts on the transverse momenta, the rapidities  and the  $\Delta R(\gamma_1,\gamma_2)$ distance of the two hardest isolated photons. As for $\tta$ production, NLO EW corrections are well within the total  uncertainty of $\NLOQCD$ predictions, although their relative impact is slightly larger for this process: $\sim -3 \%$ of the $\LOQCD$ prediction or equivalently $\sim -2 \%$ of the $\NLOQCD$ one. We want to stress again that only after performing an exact calculation such as the one presented here we can claim that {\it at the inclusive level} NLO EW corrections are negligible in comparison to the total QCD uncertainty (scale+PDF). 

In the case of $A_C$, the most striking difference with the case of $\tta$ is   its  absolute size, which is roundabout five times larger. On the other hand, total rates are, depending on the cuts, hundreds to thousand times smaller than for $\tta$ production. One should also not  forget that with a 100 TeV collider, rates will increase by roughly a factor fifty \cite{Maltoni:2015ena}, but the value of $A_C$ will also decrease. Indeed with higher hadronic energies the relative contribution of gluon--gluon initiated processes increases, but being completely symmetric it enters only the denominator of $A_C$ (see eq.~\eqref{eq:asym}). The same effects can be seen in $\tta$ by comparing results in Tab.~\ref{tab:tta}  with those in Ref.~\cite{Mangano:2016jyj}, which are for 100 TeV collisions.
Thus, while the measurement of $A_C$ in $\tta$ hadroproduction is achievable in the next future \cite{Bergner:2018lgm}, in the case of $\ttaa$ its feasibility still remains an open question. Nevertheless it is important to notice the impact of NLO corrections. NLO QCD corrections decrease the size of $A_C$, with the  $\NLOQCD/\LOQCD$ ratio ranging from 0.64 to 0.75 in the four phase-space cuts choices of Tab.~\ref{tab:ttaa}.  
Similarly to $\tta$ production,  $\LOQCD$ scale uncertainties are very small, but they are much larger when NLO QCD corrections are taken into account. The effect of NLO EW  is also not negligible, being the  $\NLOQCDEW/\NLOQCD$ ratio $\sim$ 0.96 for all the four phase-space cuts choices. Still, it is well within the total QCD uncertainties (scale+PDF), but it may be further  reduced by the missing $(\NLO-\NLOQCDEW)$ term. As already mentioned, we leave this calculation for future work.

\begin{figure}[!t]
\centering
\includegraphics[width=0.325\textwidth]{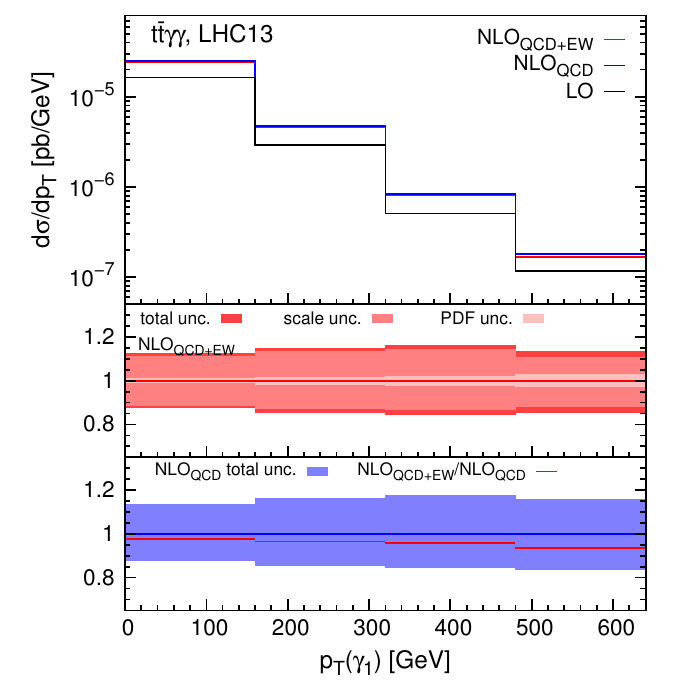}
\includegraphics[width=0.325\textwidth]{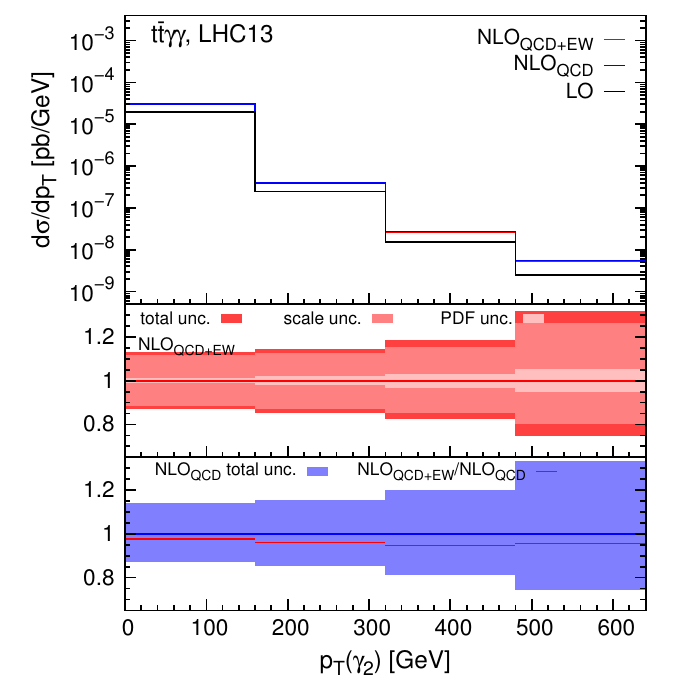}
\includegraphics[width=0.325\textwidth]{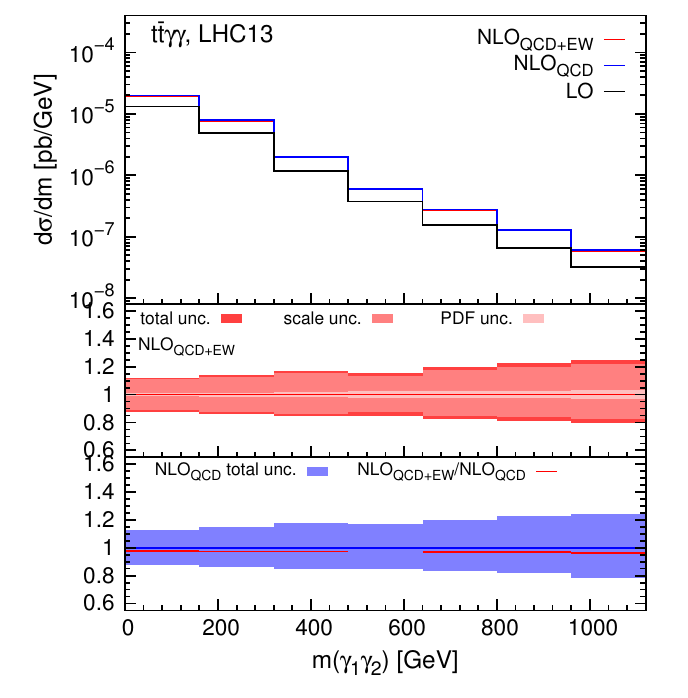}
\includegraphics[width=0.325\textwidth]{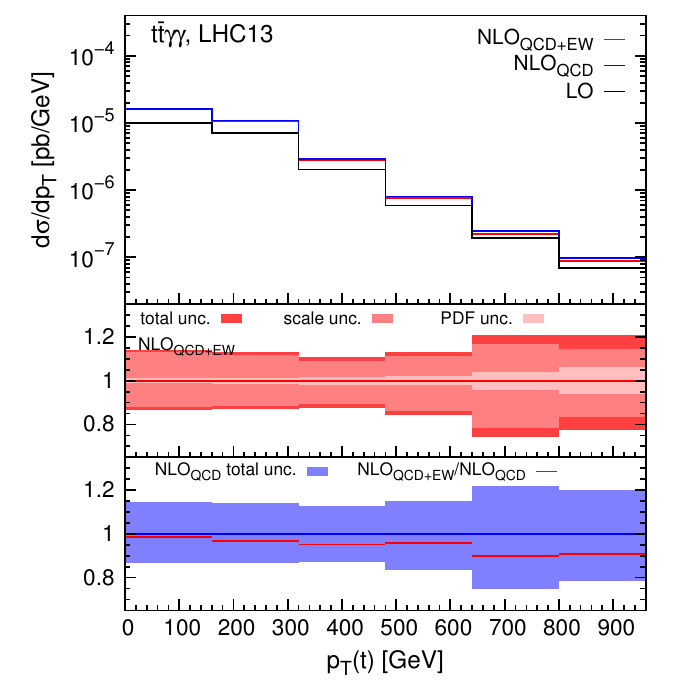}
\includegraphics[width=0.325\textwidth]{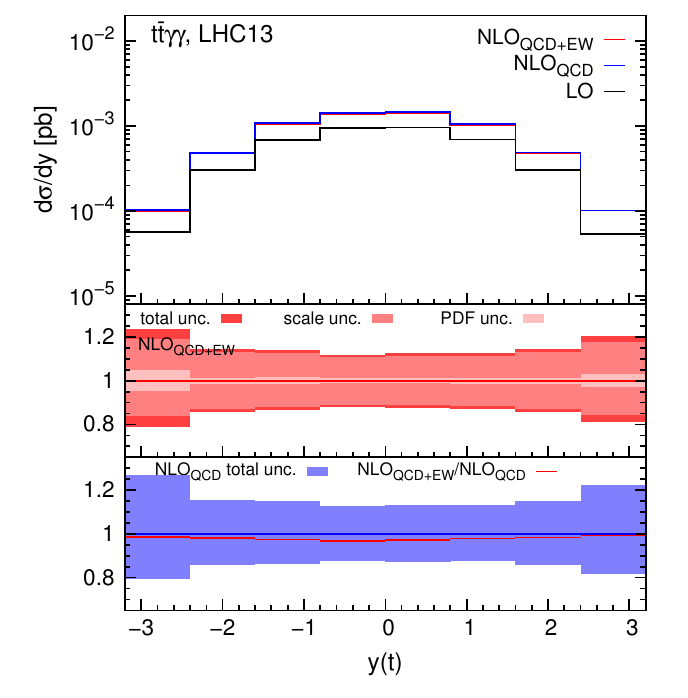}
\includegraphics[width=0.325\textwidth]{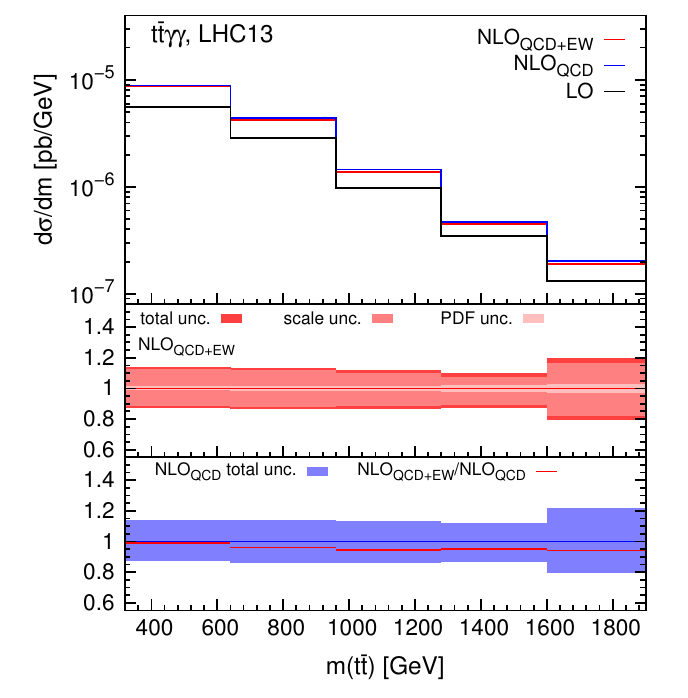}

\caption{Differential distributions for $\ttaa$ production. \label{fig:ttaa} }
\end{figure}

We now move to the case of differential distributions.  In Fig.~\ref{fig:ttaa} we show distributions for the transverse momentum of the first and second hardest isolated photons and their invariant mass, the transverse momentum and rapidity of the top quark, and the invariant mass of the top-quark pair. The layout of the plots is very similar to the one of the plots displayed in Fig.~\ref{fig:tta} and described in Sec.~\ref{sec:tta}; the only difference is that Complete-$\NLO$ predictions are not present. Most of the features described for the plots in Fig.~\ref{fig:tta} apply also for the corresponding ones presented in Fig.~\ref{fig:tta}, therefore we do not repeat them here. We notice that the largest effect of NLO EW corrections is present for the case of the $p_T(t)$ distributions, reaching in the tail almost the lower edge of total QCD uncertainties (scale+PDF). In the case of $m(\gamma_1 \gamma_2)$, which clearly was not present in Fig.~\ref{fig:tta}, effects of NLO EW corrections are well within the total QCD uncertainties.

\subsection{Single-top photon associated production: $\taj$ }
For the calculation of NLO QCD and EW corrections of single-top photon associated hadroproduction ($\taj$), we closely follow the approach of Ref.~\cite{Pagani:2020mov}, where the same kind of calculation has been performed for the single-top and $H$ or $Z$ boson associated hadroproduction. For the first time we provide $\NLOQCDEW$ predictions for $\taj$ production, together with an estimate of the flavour-scheme uncertainties, based on the procedure that has been presented,  motivated  and explained in details in Ref.~\cite{Pagani:2020mov}. Here we will not repeat the details; we invite the interested reader to look for them in Ref.~\cite{Pagani:2020mov}.  As for any other process, NLO QCD corrections to $\taj$ production can be calculated since a few years ago in a completely automated way via the {\sc\small MadGraph5\_aMC@NLO} framework \cite{Alwall:2014hca}. On the other hand, at least to the best of our knowledge, so far  a dedicated study of $\taj$ production has never been performed even at NLO QCD accuracy.\footnote{NLO QCD corrections have been calculated for this process with top-quark flavour-changing neutral interactions \cite{Zhang:2011gh, Degrande:2014tta}, where the final state is exactly $t \gamma$ without an additional jet.} Therefore, results in this section are new not only for what concerns NLO EW corrections but also at NLO QCD accuracy. We remind the reader that the CMS collaboration has already found the evidence for $\taj$ production \cite{Sirunyan:2018bsr}, and searches in the context of flavour-changing neutral currents have been performed for this process by the ATLAS collaboration \cite{Aad:2019pxo}.

At variance with $\tta$ and $\ttaa$ production,  according to eq.~\eqref{eq:blobs_NLO_general}, $k=1$ for $\taj$ production and therefore at LO only the $\LOQCD$, also denoted  $\LO_1$, contribution is present and at NLO only the NLO EW and NLO QCD corrections are present, $\NLO_1$ and $\NLO_2$ in our notation. This also means that the Complete-NLO and the NLO QCD+EW predictions coincide ($\NLO=\NLOQCDEW$). However, again at variance with $\tta$ and $\ttaa$ production,  since $n_\gamma=1$ and $\LO_1\propto \alpha^3 $, the use of the mixed scheme in principle allows for both the cases $\bar \alpha=\alphaz$ and $\bar \alpha=\alphaGmu$, as shown in \eqref{eq:NLO_rescale}. We will therefore comment more on the  choice of the value of $\bar \alpha$ for this process. 

Before moving to numerical results, we want to summarise very briefly the approach of Ref.~\cite{Pagani:2020mov}, which is used also here for estimating flavour-scheme uncertainties. First of all, it is important to note that $\taj$ process involves at LO a bottom quark in the initial state. As very well known, similarly to the case of single-top production without photons in the final state \cite{Campbell:2009ss, Maltoni:2012pa, Bothmann:2017jfv, Gao:2020ejr}, this implies that the calculation can be performed in the 4FS or 5FS. We perform our calculation in the 5FS, without selecting any particular channels ($s$-, $t$ or $tW$ associated), but we want also to take into account the uncertainty due the choice of the 5FS instead of the 4FS, for which the calculation is more cumbersome. In Ref.~\cite{Pagani:2020mov} we have motivated why the following approach should be preferred for this purpose. First, the $t$-channel only production mode is identified both in the 4FS and 5FS at NLO QCD accuracy and denoted as $\NLOQCDt^{\rm 4FS}$ and $\NLOQCDt^{\rm 5FS}$, respectively. Then, the scale uncertainties for these two quantities are evaluated via the nine-point independent variation of the renormalisation and factorisation scales, around a common central value. Next, a combined scale+flavour uncertainty band is identified as the envelope of the previous two and denoted as $\rm 5FS_{\rm 4-5}^{\rm scale}$, with the central value equal to the one in the 5FS. Finally the relative upper and lower uncertainty induced by the $\rm 5FS_{\rm 4-5}^{\rm scale}$ is then propagated to the entire $\NLOQCD$ and $\NLOQCDEW$ prediction, without selecting the $t$-channel only. All the motivations for this approach, can be found in Ref.~\cite{Pagani:2020mov}, where all the arguments underlying this procedure do not depend on the presence of the $Z$ or Higgs boson in the final state, which can therefore be substituted with the photon.

  \subsubsection{Numerical Results}

\begin{table}[!t]
\renewcommand{\arraystretch}{2.5}
\scriptsize
\begin{center}
\begin{tabular}{c c c | c c}
\hline
\hline
\multicolumn{5}{c}{\normalsize $\taj$ } \\
\hline
Accuracy & Channel & FS & Inclusive [fb] & Fiducial [fb] \\
\hline
\multirow{3}{*}{$\NLOQCD$} & \multirow{3}{*}{$t$-ch.} & 4FS & $780(1)_{-40.53 (-5.2  \%)}^{+32.37 (+4.1  \%)}~_{-3.52 (-0.5  \%)}^{+3.52 (+0.5  \%)}$ & $586(1)_{-29.29 (-5.0  \%)}^{+20.22 (+3.4  \%)}~_{-2.68 (-0.5  \%)}^{+2.68 (+0.5  \%)}$ \\
 &  & 5FS & $806(2)_{-17.22 (-2.1  \%)}^{+57.13 (+7.1  \%)}~_{-3.64 (-0.5  \%)}^{+3.64 (+0.5  \%)}$ & $599(1)_{-21.94 (-3.7  \%)}^{+55.02 (+9.2  \%)}~_{-2.84 (-0.5  \%)}^{+2.84 (+0.5  \%)}$ \\
 &  & 5FS$_{\rm 4-5}^{\rm scale}$ & $806(2)_{-66.07 (-8.2  \%)}^{+57.13 (+7.1  \%)}~_{-3.64 (-0.5  \%)}^{+3.64 (+0.5  \%)}$ & $599(1)_{-42.59 (-7.1  \%)}^{+55.02 (+9.2  \%)}~_{-2.84 (-0.5  \%)}^{+2.84 (+0.5  \%)}$  \\
\hline
\multirow{2}{*}{$\NLOQCD$} & \multirow{2}{2cm}{ \centering $t$-ch., $s$-ch., $tW_h$} & 5FS & $900(2)_{-36.26 (-4.0  \%)}^{+52.05 (+5.8  \%)}~_{-4.76 (-0.5  \%)}^{+4.76 (+0.5  \%)}$ & $677(2)_{-22.67 (-3.3  \%)}^{+51.29 (+7.6  \%)}~_{-3.74 (-0.6  \%)}^{+3.74 (+0.6  \%)}$ \\
 &  & 5FS$_{\rm 4-5}^{\rm scale}$ & $900(2)_{-73.78 (-8.2  \%)}^{+63.80 (+7.1  \%)}~_{-4.76 (-0.5  \%)}^{+4.76 (+0.5  \%)}$ & $677(2)_{-48.10 (-7.1  \%)}^{+62.14 (+9.2  \%)}~_{-3.74 (-0.6  \%)}^{+3.74 (+0.6  \%)}$ \\
\multirow{2}{*}{$\NLOQCDEW$} & \multirow{2}{2cm}{ \centering $t$-ch., $s$-ch., $tW_h$} & 5FS & $875(2)_{-33.13 (-3.8  \%)}^{+55.18 (+6.3  \%)}~_{-4.64 (-0.5  \%)}^{+4.64 (+0.5  \%)}$ & $657(2)_{-23.60 (-3.6  \%)}^{+53.54 (+8.1  \%)}~_{-3.65 (-0.6  \%)}^{+3.65 (+0.6  \%)}$ \\
 &  & 5FS$_{\rm 4-5}^{\rm scale}$ & $875(2)_{-71.77 (-8.2  \%)}^{+62.06 (+7.1  \%)}~_{-4.64 (-0.5  \%)}^{+4.64 (+0.5  \%)}$ & $657(2)_{-46.71 (-7.1  \%)}^{+60.34 (+9.2  \%)}~_{-3.65 (-0.6  \%)}^{+3.65 (+0.6  \%)}$ \\
 \hline
 \hline
\end{tabular}
\end{center}
\caption{Cross section for $t\gamma j$ production. The uncertainties are respectively the (flavour+)scale and the PDF ones in the form: $\pm$ absolute size ($\pm$ relative size). The first number in parentheses after the central value is the absolute statistical error.}
\label{tab:taj}  
\end{table} 
 
 For the definition of the phase-space cuts  we follow the analysis  performed by the  CMS collaboration \cite{Sirunyan:2018bsr}, which has led to the evidence for $\taj$ production in proton--proton collisions. Events are required to satisfy the following cuts:  
\begin{enumerate}
\item Exactly one isolated photon with $p_T(\gamma)>25$ GeV and $|\eta(\gamma)|<1.44$,
\item At least one jet with $p_T(j)>40$ GeV and $|\eta(j)|<4.7$,
\item Jet-photon separation $\Delta R(\gamma, j)>0.5$, where $j$ stands for all the jets in the event.
\end{enumerate}
Based on this we define two phase-space regions: Inclusive (only the first cut applied) and Fiducial (all cuts applied).

In Tab.~\ref{tab:taj} we report Inclusive and Fiducial results for different approximations. In the upper half of the table there are results at NLO QCD accuracy in the 4FS and 5FS for the $t$-channel mode only, together with the 5FS$_{\rm 4-5}^{\rm scale}$ prediction, whose definition has been introduced before in this section. In the lower part of the table there are results obtained without selecting the $t$-channel only, for both $\NLOQCD$ and $\NLOQCDEW$ predictions. In both cases we display the pure 5FS and the 5FS$_{\rm 4-5}^{\rm scale}$ prediction, which is derived via the procedure introduced in the previous section and based on Ref.~\cite{Pagani:2020mov}. The predictions dubbed as 5FS$_{\rm 4-5}^{\rm scale}$, including all channels and flavour+scale uncertainties, are the most precise and reliable, especially the one at $\NLOQCDEW$ accuracy, which taking into account both NLO QCD and EW corrections has to be considered as our best prediction  for $\taj$ production. The label $tW_h$ in the table refers to those diagrams consisting of $tW\gamma$ associated production with subsequent $W$ decay into quarks ($h$ for hadronic), which appear both via NLO QCD and EW corrections.

First of all, by comparing results in the upper and lower half of Tab.~\ref{tab:taj}, it is  evident how the sum of the contributions of the $s$-channel and $tW_h$ modes  exceeds the total uncertainty of the $t$-channel alone. Thus, these two contributions cannot be ignored in the comparisons between data and the SM predictions.
Then, as expected, for both the Inclusive and Fiducial results,  at NLO QCD accuracy 5FS$_{\rm 4-5}^{\rm scale}$ predictions have  larger uncertainties than  the corresponding 4FS and 5FS results. For both cuts, the NLO EW corrections are $\sim -3\%$ of the $\NLOQCD$ predictions, therefore well within the  5FS$_{\rm 4-5}^{\rm scale}$ uncertainty. On the other hand, we notice that in the pure 5FS the lower edge of the $\NLOQCD$ band would be much closer to the $\NLOQCDEW$ central prediction, for both the phase-space cuts. This fact supports the relevance of employing the 5FS$_{\rm 4-5}^{\rm scale}$ approach for obtaining reliable results. The relevance of the 5FS$_{\rm 4-5}^{\rm scale}$ approach and the importance of the NLO EW corrections can be better appreciated with differential distributions, which we are going to describe in the following.

\begin{figure}[!t]
\centering
\includegraphics[width=0.325\textwidth]{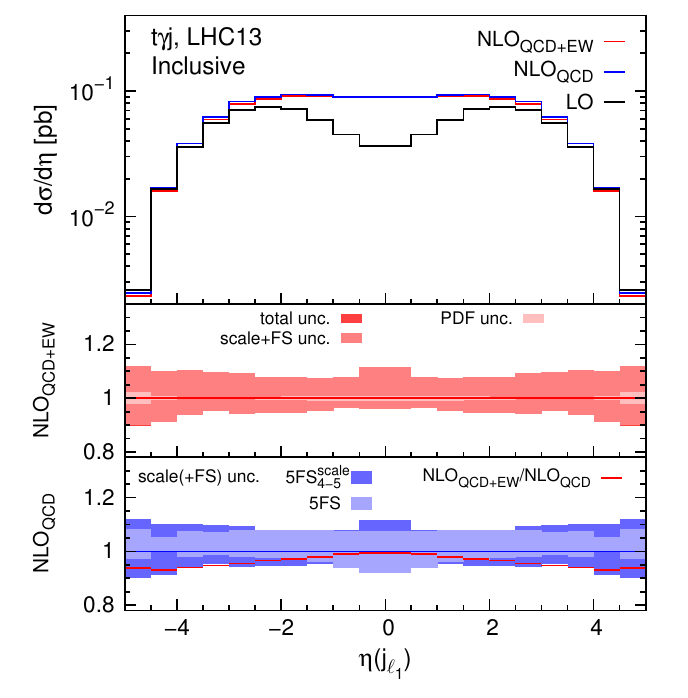}
\includegraphics[width=0.325\textwidth]{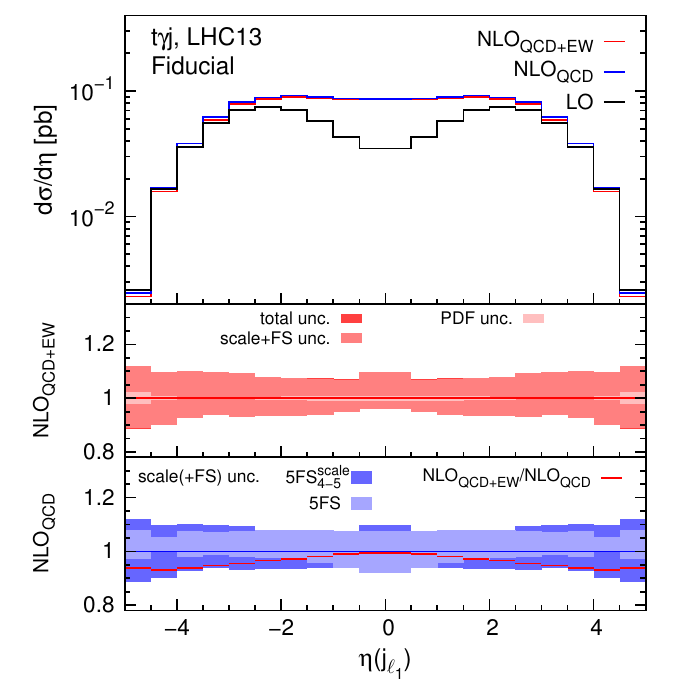} \\
\includegraphics[width=0.325\textwidth]{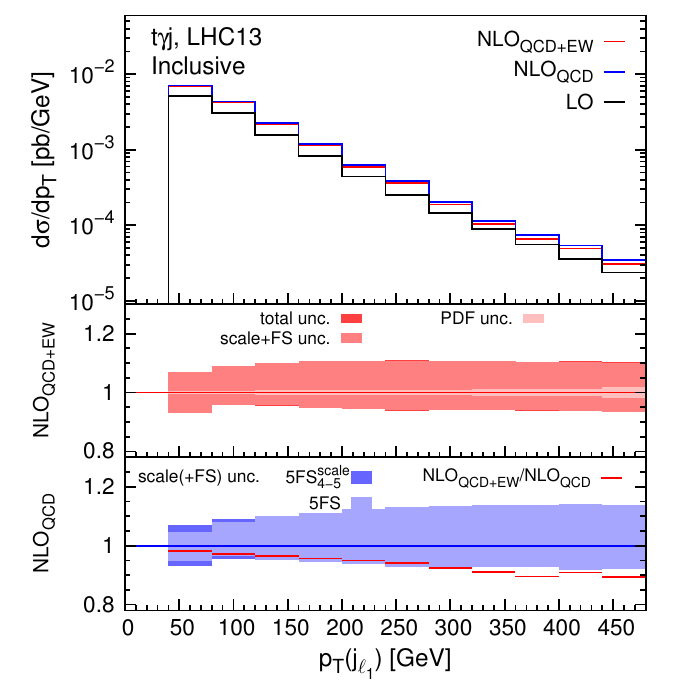}
\includegraphics[width=0.325\textwidth]{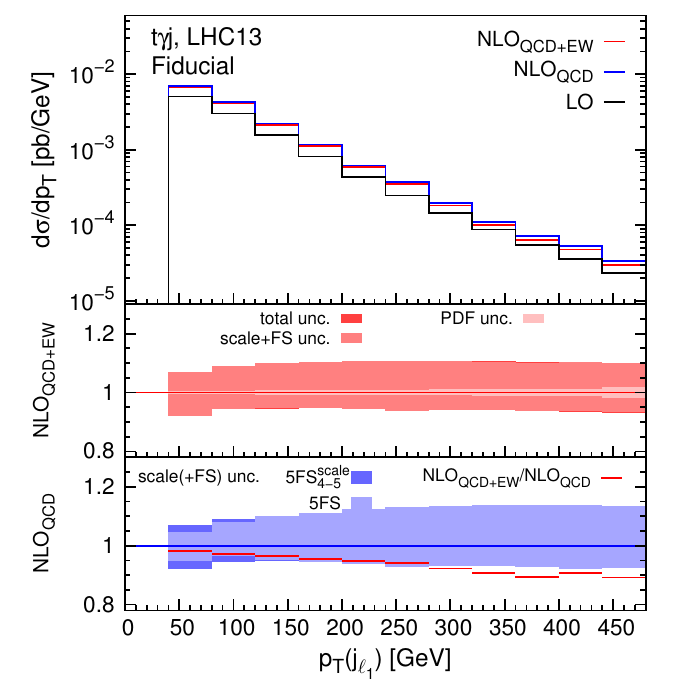} \\
\includegraphics[width=0.325\textwidth]{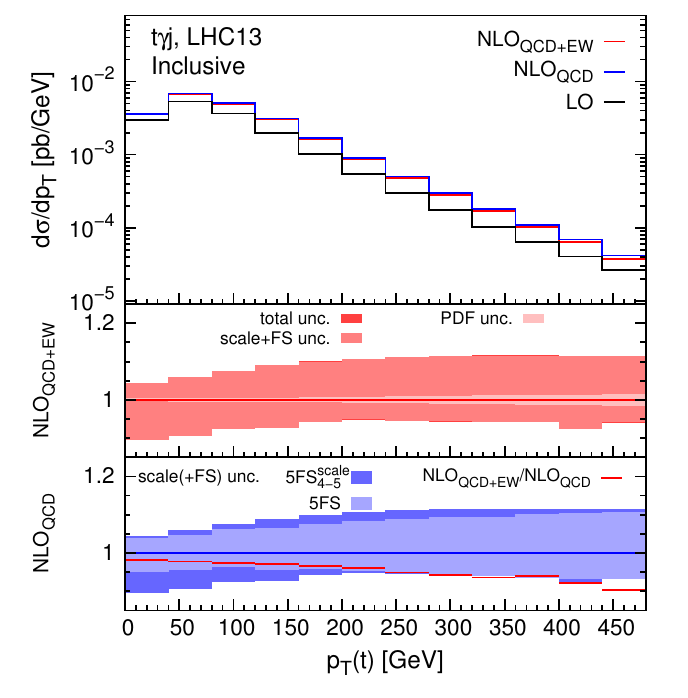}
\includegraphics[width=0.325\textwidth]{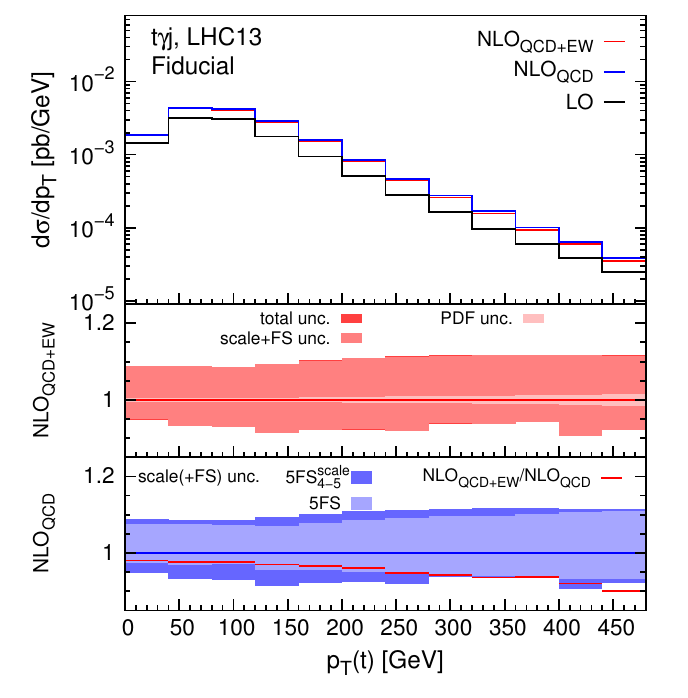} \\
\includegraphics[width=0.325\textwidth]{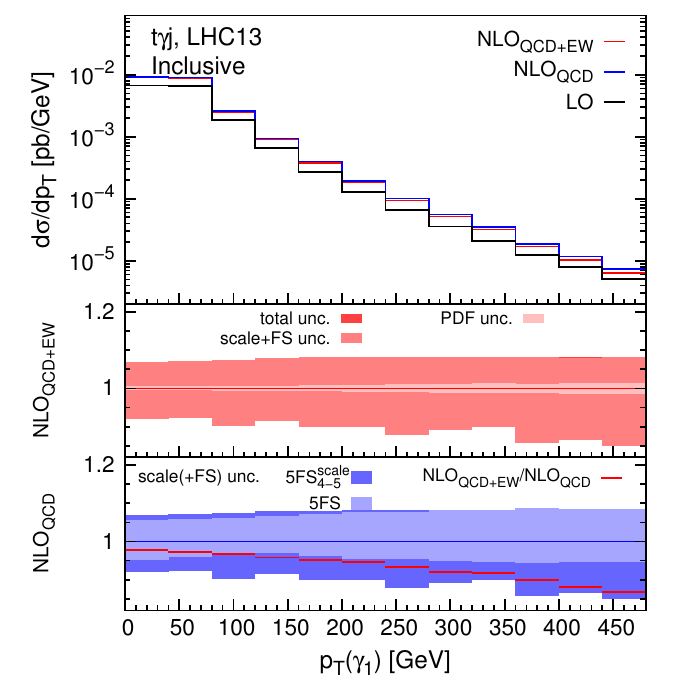}
\includegraphics[width=0.325\textwidth]{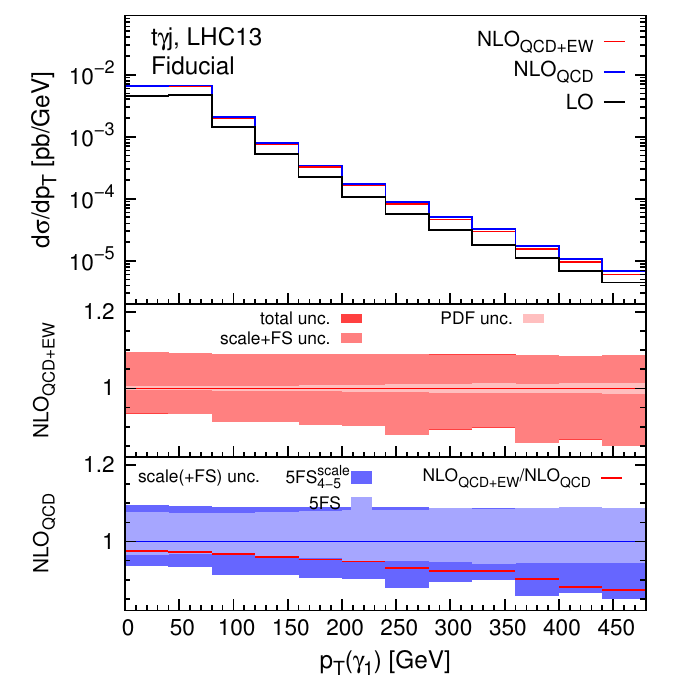}
\caption{Differential distributions for $\taj$ production.}
\label{fig:tja}
\end{figure}

In Fig.~\ref{fig:tja} we show differential distributions for $\taj$ production, without selecting the  $t$-channel. In particular, we show the pseudorapidity and transverse-momentum distributions of the hardest light-jet ($j_{l_1}$), and the transverse momentum of the top (anti)quark and  hardest isolated-photon. The plot on the left are obtained with the Inclusive cuts, while those on the right with the Fiducial one. For each plot we show in the main panel the central value of the LO, $\NLOQCD$ and $\NLOQCDEW$ predictions in the 5FS. In the first inset we separately show the relative scale+flavour and PDF uncertainty of the $\NLOQCDEW$ prediction together with their sum in quadrature, the total uncertainty. In the last inset we show the 5FS scale and 5FS$_{\rm 4-5}^{\rm scale}$ scale+flavour relative uncertainties for the NLO QCD prediction, together with the $\NLOQCDEW/\NLOQCD$ ratio. 

First of all we see that plots for Inclusive and Fiducial cuts are almost identical, besides their normalisations. The only exception is the threshold region for the $p_T(t)$ distribution. Thus, the following considerations apply to both cases. The large difference between LO and $\NLOQCD$ or $\NLOQCDEW$ predictions in the central region of the  $\eta(j_{l_1})$ distributions is due to the opening of the $tW_h$ channel via the NLO corrections, which, as explained in Refs.~\cite{Frederix:2019ubd, Pagani:2020mov}, is not enhanced for large $\eta(j_{l_1})$ values and therefore populates the central region of this distribution. In the peripheral region, if we did not take into account flavour uncertainties, namely in the 5FS, NLO EW corrections would be larger than the QCD scale-uncertainty band; only with the 5FS$_{\rm 4-5}^{\rm scale}$ approach are within it. The same argument applies to the tail of the $p_T(\gamma_1)$ distribution, where although NLO EW corrections reach the size of $\sim-15\%$ of the $\NLOQCD$ prediction,\footnote{For this process,  the large size of the EW corrections in the tail is partially due to the fact that we require exactly one isolated photon. Indeed, since  part of the photon radiation with $p_T>25~\gev$ is vetoed, an additional negative correction that grows in absolute size in the tail is present.} they are still within  5FS$_{\rm 4-5}^{\rm scale}$ total uncertainty. The situation is instead different in the tail of the $p_T(j_{l_1})$ and $p_T(t)$ distributions, where NLO EW corrections are larger than 5FS$_{\rm 4-5}^{\rm scale}$ uncertainties, which on the other hand almost overlap with the 5FS ones. 

In conclusion, no sizeable differences have been observed between results for the Inclusive and Fiducial regions, besides the total rates, and the 5FS$_{\rm 4-5}^{\rm scale}$ approach should be preferred both for total and differential rates. Only following this approach, NLO EW corrections are in general within the total uncertainty, but also in this case exceptions are present in the tail of distributions. We also have compared results obtained with $\bar \alpha=\alphaz$ and $\bar\alpha=\alphaGmu$ in order to assess how large is the numerical impact of the choice of the value of $\bar\alpha$, where the latter choice is superior from a formal point of view.  As expected, even in the tail of the $p_T(\gamma_1)$ distribution, where corrections have been found to be sizeable, the choice of the value for $\bar \alpha$ had an impact below the percent level. In general, results obtained via the two different choices of $\bar \alpha$ have been found compatible within their numerical accuracy.
 
\subsection{Top-quark decay involving photons: $\tadlep$ and $\tadhad$}
\label{sec:decay}

\begin{table}[!t]
\renewcommand{\arraystretch}{2.25}
\begin{center}
\begin{tabular}{c c | c c }
\hline
\hline
\multicolumn{3}{c}{$\Gamma_t$ [MeV]} \\
\hline
 Order & $\tadhad$ & $\tadlep$ \\
\hline
  LO & $4.433(2)$ & $2.870(2)$ \\
   NLO$_{\rm QCD}$ & $3.52(4)^{+2.65 \%}_{-3.22 \%}$ & $2.550(6)^{+1.39 \%}_{-1.68 \%}$ \\
   NLO$_{\rm QCD+EW}$ & $3.50(4)^{+2.68 \%}_{-3.25 \%}$ & $2.559(9)^{+1.38 \%}_{-1.68 \%}$ \\
\hline
\hline
\end{tabular}
\end{center}
\caption{Top-quark hadronic ($\tadhad$) and leptonic  ($\tadlep$) partial decay widths. The leptonic case includes all the three leptons $e,\mu$ and $\tau$. \label{tab:decay}}  
\end{table}

As discussed in {\it e.g.}, Refs.~\cite{Melnikov:2011ta,Bevilacqua:2019quz} for the case of $\tta$ production, when top quark decays are taken into account, the contribution of photons radiated via the top decay is sizeable. The predictions for $\tta$, $\ttaa$, and $\taj$ that we have discussed in the previous sections do not include this contribution, being the top quark stable. For each of the previous processes, if top decays were considered,  an important contribution would be given by the same process without one isolated photon in the final state ($t \bar t$, $\tta$ and $t j$ respectively) and the subsequent $t\TO b W \gamma$ decay for one of the top quarks. On the other hand, the focus of this work is the calculation of EW corrections. We have shown that, besides in the tails of the distributions, NLO EW corrections are in general within the QCD uncertainties for the case with photon emitted by the hard process. In NWA, the case of the photons emitted from the top decay depends on two factors. First, the NLO EW corrections to the $t \bar t$, $\tta$ and $t j$ production processes. Second, the NLO EW corrections to the top-quark decay $t\TO b W \gamma$. The former are documented in the literature \cite{Czakon:2017wor, Frederix:2018nkq} or discussed in this work in the case of $\tta$. The latter are calculated for the first time in this section.

As already mentioned, we calculate the NLO QCD+EW predictions for the leptonic and hadronic top-quark decays $\tadlep$ and $\tadhad$. In the case of $\tadlep$ we actually select the channel $t \rightarrow \mu^+ \nu_\mu b \gamma$ for the calculation, although all the others leptonic channels are equivalent, assuming massless $\tau$ leptons. The case with top antiquarks gives  the same results.
 For this process, according to eq.~\eqref{eq:blobs_NLO_general}, $k=1$ and therefore only NLO EW and NLO QCD corrections are present ($\NLO_1$ and $\NLO_2$ in our notation), where the $\LO_1$ is proportional to $\alpha^3$. 
 
For this calculation part of the settings listed in Sec.~\ref{sec:setup} are modified. First, the central value of the renormalisation scale is set to $\mt$, then the Frixione isolation algorithm is adapted to the case of a decay process in its rest frame. The isolation is performed by using, instead of  $R_0(\gamma)=0.4$ and $p^{\rm min}_T(\gamma)>25~\gev$ like in  \eqref{eq:FP}, the parameters: $E^{\rm min}(\gamma)>25~\gev$ and $\theta_0(\gamma)=0.1$, where the separation of the photons and hadronic or electromagnetic activities is performed by looking at the separation angle.

In Tab.~\ref{tab:decay} we report LO, $\NLOQCD$ and $\NLOQCDEW$ predictions for the partial widths of $\tadhad$ and $\tadlep$ decays.
The NLO QCD corrections reduce the LO prediction by $-23\%$ and $-11\%$ for the hadronic and leptonic case, respectively. In both cases, NLO QCD scale uncertainties are only a few percents of the absolute value. NLO EW corrections are in both cases smaller than 1\% of the LO prediction. 
\begin{figure}[!t]
\centering
\includegraphics[width=0.475\textwidth]{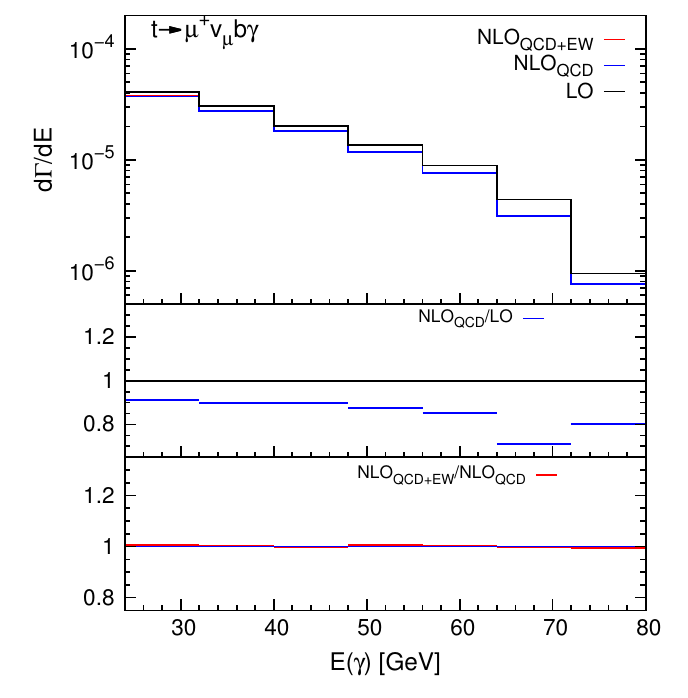}
\includegraphics[width=0.475\textwidth]{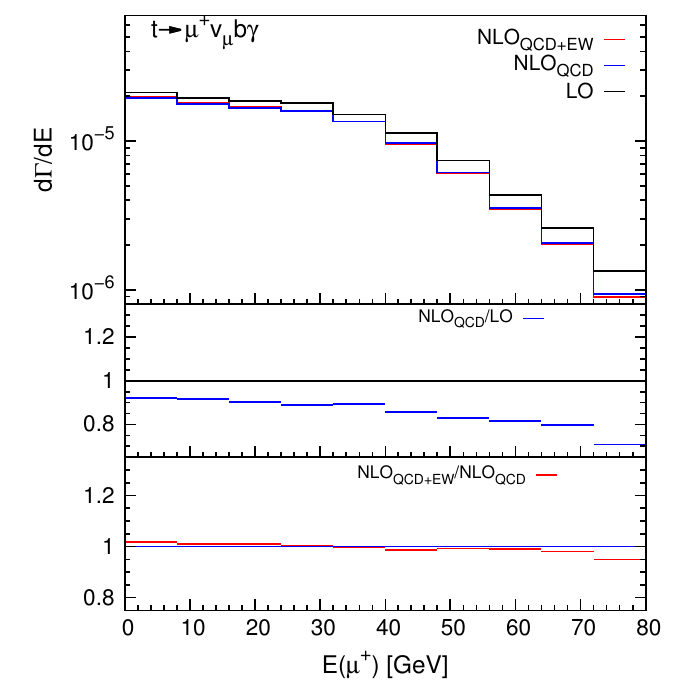}
\caption{Differential distributions for leptonic top-quark    ($\tadlep$) partial decays, with $\ell=\mu$. \label{fig:decay}} 
\end{figure}
\noindent

In Fig.~\ref{fig:decay} we show for the leptonic case the energy spectrum of the hardest photon, $E(\gamma_1)$, and of the lepton, $E(\ell^+)$. In the main panel we show LO, $\NLOQCD$ and $\NLOQCDEW$ predictions, in the first inset we show the $\NLOQCD/\LO$ ratio and in the second inset the $\NLOQCDEW/\NLOQCD$ ratio. As can be seen, while the relative impact of NLO QCD corrections varies a lot in both distributions, the NLO EW corrections remain at or below the percent level in the full spectrum, with the exception of the tail of the distribution in the case of $E(\ell^+)$. Needless to say, although  the choice $\bar \alpha = \alphaGmu$ in \eqref{eq:NLO_rescale} is formally superior, in practice the choice of the value of $\bar \alpha  $ is completely negligible.

The results obtained for the $\tadhad$ and $\tadlep$ decays point to the fact that, when looking at $\tta$, $\ttaa$ or $\taj$ production, although the contribution of photons emitted after the top decay is in general sizeable, the size of the NLO EW corrections to the top-quark decay in association with photons is negligible. A study of the NLO EW corrections for the complete final state $W^+bW^- \bar b \gamma$ including $W$ decays  in NWA or with full off-shell effects, as already done for NLO QCD in respectively Ref.~\cite{Melnikov:2011ta} and Ref.~\cite{Bevilacqua:2018woc}, is definitely worth to be considered, but beyond the scope of this paper. The same applies to the $W^+bW^- \bar b \gamma \gamma$ and $Wb j \gamma$ final states.

 \section{Conclusions}\label{sec:conclusions}
In this paper we have calculated for the first time:

\begin{itemize}
\item the Complete-NLO predictions for top-quark pair production in association with at least one photon ($\tta$),
\item the NLO QCD+EW corrections for top-quark pair production in association with at least two photons ($\ttaa$),
\item the NLO QCD+EW corrections for single-top production in association with one photon ($\taj$), together with a 4FS and 5FS comparison,
\item the NLO QCD+EW corrections for leptonic ($\tadlep)  $ and hadronic ($\tadhad$) decays.
\end{itemize} 

In the case of cross sections, we find that EW corrections are in general within QCD uncertainties. For $\taj$ production, that is true  only if the uncertainty due to the flavour scheme is taken into account. Moreover, for this process, in the tail of the distributions EW corrections are sizeable and of the same size of (or larger than) QCD uncertainties. We also have analysed the top-quark charge asymmetry $A_C$ for $\tta$ and $\ttaa$ production and found sizeable effects for NLO QCD and NLO EW corrections and as well for subleading NLO orders.
Therefore, unlike other processes involving top quarks ($t \bar t W$ and $t \bar t t \bar t$), EW corrections are under control for this class of processes  and have a size that is of the order estimated  from the naive  $\alphas$ and $\alpha$ power counting. We want to stress that this conclusion can be drawn only after having performed an exact calculation of NLO corrections, as done here in this work.  

All these calculations have been performed in a completely automated approach via the {\mglong} framework, without any dedicated customisation for the processes considered.  In order to achieve this, we have extended the capabilities of the {\sc\small MadGraph5\-\_aMC@NLO} framework,  enabling the calculation of Complete-NLO predictions for processes with isolated photons in the final state. In this work we have discussed the technical details of the implementation, which involves a mixed EW renormalisation scheme ($\alphaz$ and $G_\mu$ or $\alpha(\mz)$) for this class of processes. We have also discussed the issues related to the choice of the numerical value of $\alpha$ in the $\ord(\alpha)$ corrections and the subtleties related to this aspect in the context of automated calculations.

 \section*{Acknowledgements}

 We are grateful to   the developers of {\sc MadGraph5\_aMC@NLO} for the long-standing collaboration and for discussions.    
   The work of D.P.~is supported by the Deutsche Forschungsgemeinschaft (DFG) under Germany's Excellence Strategy - EXC 2121 ``Quantum Universe'' - 390833306. H.-S.S.~is supported by the European Union's Horizon 2020 research and innovation programme under the grant agreement No.824093 in order to contribute to the EU Virtual Access ``NLOAccess'', the French ANR under the grant ANR-20-CE31-0015 (``PrecisOnium''), and the CNRS IEA under the grant agreement No.205210 (``GlueGraph"). The work of I.T.~is supported by the Swedish Research Council under contract number 2016-05996 and the MorePheno ERC grant agreement under number 668679. Computational resources to I.T.~have been provided by the Consortium des \'Equipements de Calcul Intensif (C\'ECI), funded by the Fonds de la Recherche Scientifique de Belgique (F.R.S.-FNRS) under Grant No.~2.5020.11 and by the Walloon Region.
M.Z.~is supported by the ``Programma per Giovani Ricercatori Rita Levi Montalcini'' granted by the Italian Ministero dell'Universit\`a e della Ricerca (MUR).

\bibliographystyle{utphys}

 \bibliography{draft} 
 
\end{document}